\documentclass{jfm}

\usepackage{graphicx}
\usepackage{newtxtext}
\usepackage{newtxmath}
\usepackage{natbib}
\usepackage{hyperref}
\usepackage{amsmath}
\usepackage[x11names]{xcolor}
\usepackage{subcaption}
\usepackage{multirow}

\hypersetup{
    colorlinks = true,
    urlcolor   = blue,
    citecolor  = black,
}

\newcommand{\RomanNumeralCaps}[1]
\linenumbers

\title{State-dependent convergence of Galerkin-based reduced-order models for Couette flow}

\author{Zilin Zong\aff{1} \corresp{\email{zilin.zong22@imperial.ac.uk}}, Igor Maia \aff{2}, André Cavalieri \aff{2} \and Yongyun Hwang\aff{1}}

\affiliation{\aff{1}Department of Aeronautics, Imperial College London, South Kensington, London SW7 2AZ, UK
\aff{2} Instituto Tecnológico de Aeronáutica, São José dos Campos, 12228-900, Brazil}

\begin{document}

\maketitle

\begin{abstract}

In this study, we explore the effect of basis functions on the performance and convergence of the Galerkin projection-based reduced-order model (ROM) in the minimal flow unit of Couette flow. POD (proper orthogonal decomposition) modes obtained from direct numerical simulation and controllability and balanced truncation modes from the linearised Navier-Stokes equations (LNSE) with different base flows (laminar base flow and turbulent mean flow) and an eddy viscosity model are considered. In the neighbourhood of the laminar base state, the ROMs based on the modes from the LNSE with the laminar base flow and molecular viscosity are found to perform very well as they are able to capture the linear stability of the laminar base flow for each plane Fourier component only with a single degree of freedom. In particular, the ROM based on the balanced truncation modes models the linear dynamics involving transient growth around the laminar base flow most effectively, consistent with previous studies. In contrast, for turbulent state, the ROM based on POD modes is found to reproduce its statistics and coherent dynamics most effectively. The ROMs based on the modes from the LNSE with turbulent mean flow and an eddy viscosity model performs better compared to any other ROMs using the modes from the LNSE. These observations suggest that the performance and convergence of a ROM are highly state-dependent. In particular, this state dependence is strongly correlated with the information and dynamics that each of the basis functions contains. Discussions supporting these observations are also provided in relation to the flow physics involved and the form of coherent structures in Couette flow.  
\end{abstract}



\section{Introduction}
\label{sec:introduction}

Modelling a turbulent flow often involves a large number of degrees of freedom, especially at high Reynolds numbers. One of the most fundamental challenges in this modelling process is to have an accurate description of the spatio-temporal evolution of flow fields while reducing the degrees of freedom (DoF) as much as possible. 
An approach that has been pursued is the development of reduced-order models (ROM) based on the concepts of dynamical systems. In this approach, the development of a finite-dimensional dynamical system with much reduced DoF is aimed at \citep[see review by][]{c_rowley_control_review}. It has seen success in flow modelling especially for identification of the dominant coherent structures and related physical processes \cite[e.g.][]{sirovich, Aubry_Holmes_Lumley_Stone_1988_JFM,Holmes_Lumley_Berkooz_1996,10.1063/1.869185,Moehlis_2004_1,ROM_couett_lifetime.6.034610,Cavalieri_Rempel_Nogueira_2022,McCormack_Cavalieri_Hwang_2024} and flow control applications in transitional flows \citep[e.g][and many others]{ROWLEY2004_stability,rowley_2,ilak/1.2840197,sipp_1,c_rowley_control_review, maia2024turbulence}.  

Perhaps, one of the most well-studied approaches in reduced-order modelling is based on a Galerkin projection of the equations of motion using the proper orthogonal decomposition (POD) modes \citep{sirovich, Aubry_Holmes_Lumley_Stone_1988_JFM,Holmes_Lumley_Berkooz_1996,TR_smith,NOACK_AFANASIEV_MORZYŃSKI_TADMOR_THIELE_2003,Smith2005}. This method first obtains a set of orthogonal basis functions that best represent the given flow in terms of its energy content (POD modes) and subsequently projects the equations of motion onto the space spanned by these basis functions. With suitable truncation in the number of POD modes, a finite-dimensional representation of the original fluid flow is obtained in the form of a dynamical system. An important challenge in this type of approach is often associated with truncation in the number of POD modes. The exclusion of higher-order POD modes, which often represent small-scale motions with a lower energy content, causes an issue in modelling dynamically important, but not energetic, flow structures involved in complex nonlinear processes, such as turbulent energy cascade and dissipation. This greatly affects the accuracy of the ROM and sometimes causes numerical instabilities. To address this limitation, additional modelling efforts have been made, such as incorporation of eddy viscosity, to improve the accuracy and stability of ROM \citep[see e.g.][]{Rempfer_Fasel_1994a, Rempfer_Fasel_1994, TR_smith,Smith2005,Osth_2014,Chua_Khoo_Chan_Hwang_2022}.

POD modes are, however, not the only possible orthogonal basis functions that can be used for the design of ROMs. In principle, any set of orthogonal functions satisfying the boundary conditions may be used to approximate the Navier-Stokes equations into a finite-dimensional dynamical system due to its fast (often exponential) converging nature, and this is the theoretical basis underpinning the spectral discretisation \cite[e.g.][]{Canuto88}. Indeed, some ROMs that aim to model the key underlying physical processes are based on simple orthogonal functions, such as trigonometric functions \cite[e.g.][]{10.1063/1.869185,Moehlis_2004_1,ROM_couett_lifetime.6.034610}. Furthermore, some recent studies have used eigenfunctions of the controllability Gramian operator of the linearised Navier-Stokes equations: i.e. controllability modes \cite[][]{PhysRevFluids.7.L102601,McCormack_Cavalieri_Hwang_2024,maia2024GQL}. The controllability Gramian operator is the covariance of the stochastic response of a given linear dynamical system. Therefore, the resulting modes are the POD modes of the flow fields from the linearised Navier-Stokes equations subject to a stochastic forcing. If the equations are used to model a turbulent flow, the corresponding controllability modes may well be a good candidate for the basis functions to design a ROM; such operator-based modes do not require large data sets, and may thus lead to ROMs directly from governing equation. 
POD modes and the controllability modes are the basis functions that represent the most energetic structures in the given flow fields, but they do not necessarily capture all dynamically important structures effectively. This is particularly true when energy-containing flow structures are initiated by some non-energetic structures with very different shapes. Balanced truncation is a model reduction method that accounts for this issue from a viewpoint of control theory for linear dynamical systems \citep{mooreBPOD,zhoukemin}. Furthermore, an approach of approximating the associated modes (i.e. balanced truncation modes) has been proposed to deal with the high state-space dimension of the linearised Navier-Stokes equations \cite[][]{rowley_2}: balanced POD modes. Application of this type of approaches has been shown to be particularly useful for modelling the flow dynamics associated with non-normality of the linearised Navier-Stokes equations \citep{ilak/1.2840197,shervin_1}. 


The scope of the present study is to systematically develop a ROM based on Galerkin projections for turbulent flows, with the ultimate goal of reaching high Reynolds numbers. Recently, there have been an increasing number of studies to develop ROMs for turbulent flows using data-driven approaches
\citep[e.g.][and many others]{Brunton_PNAS_16,Kaszas_haller_prf22,Linot_Graham_2023}, but here we shall focus on the construction of ROMs based on Galerkin projections. Although this approach is classical and its capability in reducing the DoF may not be as good as that of the data-driven counterparts, it offers some important advantages from a physical viewpoint. First, it preserves the form of the original Navier-Stokes equations, allowing one to interpret and assess its complex chaotic dynamics in terms of the turbulent energy budget \cite[i.e. turbulence production, inter-scale and space turbulent transport, dissipation, etc; see][for example]{McCormack_Cavalieri_Hwang_2024}. Second, the Galerkin projection is robust compared to data-driven approaches, as it neither has any overfitting issues nor relies on the geometrical structure of state space. This is important, especially at high Reynolds numbers, where the degree of freedom of a suitable ROM is generally expected to be large and the geometry of state space would be too complicated to obtain useful physical insights. Finally, when properly implemented, the Galerkin projection ensures that the ROM converges to the corresponding DNS, providing a way to properly assess its performance. 

Of particular interest in the present study is understanding how the form of the basis functions affects the convergence and performance (especially with severe truncation) of a ROM around both laminar and turbulent state, in the hope of finding a suitable set of basis functions that are efficient and computationally economical to generate for a given modelling purpose. To this end, we will consider various basis functions discussed above: 1) POD, 2) controllability, and 3) balanced truncation modes. Furthermore, the controllability and balanced modes will be generated by considering a few different forms of the linearised Navier-Stokes equations. We note that the POD modes are generated by the flow field, i.e. the modelling target itself. Therefore, numerical simulations or experiments need to be carried out before reduced-order modelling, and they are often expensive. In contrast, the controllability and balanced truncation modes are generated by computing controllability and observability Gramian operators of the linearised Navier-Stokes equations. In parallel shear flows, the computational cost of these Gramian operators is modest because Fourier decomposition in the wall-parallel directions allows their size to be relatively small \cite[e.g.][]{10.1063/1.858894,JOVANOVIĆ_BAMIEH_2005,HWANG_COSSU_2010_amplification,HWANG_COSSU_2010_1}. Even in complex flows, there are a few suitable computational approaches to approximating them \citep[e.g][]{rowley_2,ilak/1.2840197}. However, in this case, each Gramian operator may well be formulated to be a function of the flow about which the linearised Navier-Stokes equations are obtained. For modelling of transition, the laminar base flow may be the obvious choice \cite[]{JOVANOVIĆ_BAMIEH_2005,rowley_2,ilak/1.2840197}, but this is not necessarily the case for turbulent flows. Although the laminar base flow can still be used for the generation of Gramian operators \cite[]{Cavalieri_Rempel_Nogueira_2022,McCormack_Cavalieri_Hwang_2024}, turbulent mean flow, which may be approximated by considering an RANS model in practice, can also be used \citep[][]{bulter_farrel_93_pof,HWANG_COSSU_2010_amplification,McKEON_SHARMA_2010_JFM}. More recently, there has been growing evidence that the linearised Navier-Stokes equations enhanced by a suitable eddy viscosity \citep[][]{Reynolds_Hussain_1972} provide a better description of coherent structures in turbulent flows. \citep[e.g.][]{hwang_mesolayer_16,Illingworth_Monty_Marusic_2018,Morra_Semeraro_Henningson_Cossu_2019,Pickering_Rigas_Schmidt_Sipp_Colonius_2021, Ivan_eddyvisc_spod}.

The emphasis of the present study will therefore be on understanding the effectiveness of different basis functions in approximating each state space (around laminar or turbulent state) using a Galerkin-based ROM. In addition to POD modes obtained from turbulent flow, several sets of controllability and balanced truncation modes with various forms of the linearised Navier-Stokes operator will be considered: for example, variations in base flow profile (laminar base flow or turbulent mean flow) and presence of the eddy viscosity model. To identify the strength and weakness of each of the basis functions in different parts of the state space, we evaluate the linear stability and transient growth of the laminar state while examining statistics (mean and fluctuations) and dynamics (time correlations) of turbulent state. The benchmark case is chosen as plane Couette flow at a relatively low Reynolds number within a minimal flow unit \citep{Jiménez_Moin_1991,Hamilton_Kim_Waleffe_1995}, a canonical parallel shear flow where the dynamics and coherent structures are well understood. 

This paper is organised as follows. In \S\ref{sec:background}, we provide a general, but concise, introduction to the basis functions to be examined: POD, controllability, and balanced truncation modes. POD modes and the linear operators for the generation of controllability and balanced truncation modes will then be introduced in \ref{sec:3}. In \S\ref{Result}, the performance and convergence of each ROM will be presented around both laminar and turbulent state, followed by detailed discussions in \S\ref{discussion}. The paper will conclude in \S\ref{conclusion}, with some perspectives.

\section{Basis functions}\label{sec:background}
We start by introducing the different sets of basis functions to be used in the present study. Throughout this section, these basis functions are denoted in matrix form, $\Phi(\boldsymbol{x}) =[ \phi_1(\boldsymbol{x})~\phi_2(\boldsymbol{x})~...]$, so that each column represents the individual basis function. If the basis functions for incompressible flows are sufficiently smooth, satisfy homogenous or periodic boundary condition and are divergence-free, a velocity fluctuation field $\mathbf{u}(\boldsymbol{x},t)$ ($\boldsymbol{x}$ and $t$ are space and time, respectively) may be written in terms of their linear combination: 
\begin{equation} 
    \boldsymbol{u}(\boldsymbol{x},t) = \sum_{i=1}^\infty a_i(t){\phi}_i(\boldsymbol{x}),
    \label{general_ROM_expansion}
\end{equation}
where $\phi_i$ is the $i$-th basis function. The reconstructed velocity field in (\ref{general_ROM_expansion}) satisfies the continuity equation and the boundary conditions by construction. If the basis functions $\phi_i(\boldsymbol{x})$ are orthogonal, the series expansion in (\ref{general_ROM_expansion}) would often converge quickly for sufficiently smooth $\boldsymbol{u}(\boldsymbol{x},t)$. If not (e.g. balanced modes), the convergence is not necessarily guaranteed in general, although such a behaviour has not been observed in the present study. 


\subsection{POD modes}
The POD seeks the basis functions $\phi$ by formulating an optimisation problem that maximises the projection of $\boldsymbol{u}$ onto $\phi$, such that 
\begin{equation}
    \max_\phi \frac{ \overline{| \langle \boldsymbol{u}, \phi  \rangle |^2} }{ \langle \phi,\phi \rangle },
    \label{POD_opti}
\end{equation}
where $\overline{(\cdot)}$ denotes ensemble average and $\langle \cdot , \cdot \rangle$ represents the inner product defined by $\langle \boldsymbol{f},\boldsymbol{g}\rangle = \int_\Omega \boldsymbol{g}^H\boldsymbol{f}~\textrm{d}\boldsymbol{x}$ in the flow domain $\Omega$, with the superscript $(\cdot)^H$ indicating complex conjugate transpose. The first-order necessary condition for the solution to (\ref{POD_opti}) yields an eigenvalue problem that takes the form of
\begin{subequations}
\begin{equation}
    \int_\Omega \overline{\boldsymbol{u}(\boldsymbol{x})\boldsymbol{u}^H(\boldsymbol{x}')} \Phi(\boldsymbol{x}') \textrm{d}\boldsymbol{x}' = \Phi(\boldsymbol{x}) \Lambda, 
    \label{POD_eig}
\end{equation}
where
\begin{equation}
    \Lambda = \textrm{diag}[\lambda_1, \lambda_2,...]. \label{POD_eig2}
\end{equation}
\end{subequations}
Here, the eigenvector $\phi_i$ in each column of $\Phi(\boldsymbol{x})$ forms the individual POD mode, and the eigenvalue $\lambda_i$, which is positive and real,  indicates the content of the averaged kinetic energy in each POD mode. The POD modes are also orthogonal: i.e. $\langle \phi_i, \phi_j \rangle = \delta_{ij}$, where $\delta_{ij}$ is the Kronecker delta and $\phi_i$ is normalised accordingly. Consequently, the POD modes form an optimal set of basis functions for reconstruction of the flow field in terms of kinetic energy. 


\subsection{Controllability modes}\label{sec:2.2}
We now assume that the velocity fluctuation field $\boldsymbol{u}$ is obtained from the following linear dynamical system driven by a stochastic body forcing: 
\begin{subeqnarray}
    \frac{\partial{\boldsymbol{q}}}{\partial t} &=& \boldsymbol{A}\boldsymbol{q} + \boldsymbol{B}\boldsymbol{f}, \label{eq:2.4a}\\
    \boldsymbol{u} &=& \boldsymbol{C}\boldsymbol{q},
    \label{general_dyn_sys_1}
\end{subeqnarray}
where $\boldsymbol{q}$, $\boldsymbol{A}$, $\boldsymbol{B}$ and $\boldsymbol{C}$ are defined later in \S\ref{sec:3.3} for plane Couette flow. If the body forcing is white in time such that $\overline{\boldsymbol{f}(t_1)\boldsymbol{f}^H(t_2)}=\boldsymbol{I}\delta(t_1-t_2)$, the covariance operator of the velocity fluctuation is obtained from
\begin{equation}
 \boldsymbol{W}_c\chi=\int_\Omega \overline{\boldsymbol{u}(\boldsymbol{x})\boldsymbol{u}^H(\boldsymbol{x}')} \chi(\boldsymbol{x}') \textrm{d}\boldsymbol{x}', 
\end{equation}
where $\chi(\boldsymbol{x})$ is a test function sharing the same dimension with $\boldsymbol{u}$ and
\begin{equation}
\boldsymbol{W}_c=\boldsymbol{C}\boldsymbol{W}_c^q\boldsymbol{C}^{\dag}\psi
\end{equation}
with
\begin{equation}
    \boldsymbol{A}\boldsymbol{W}_c^q + \boldsymbol{W}_c^q\boldsymbol{A}^{\dag} + \boldsymbol{BB}^\dag= 0. \\ 
    \label{control_lyap}
\end{equation}
Here, the superscript $(\cdot)^\dag$ denotes adjoint with respect to the inner product defined above, and $\boldsymbol{W}_c^q$ is the controllability Gramian that characterises the reachability of $\boldsymbol{q}$ in state space by a suitable action of forcing $\boldsymbol{f}$. 
Similarly to POD modes, the following eigenvalue problem produces a set of orthogonal basis functions that are ranked in terms of the most responsive velocity field $\boldsymbol{u}$ to the white-in-time forcing:
\begin{equation}
   \boldsymbol{W}_c \Phi = \Phi\Lambda.
\end{equation}
In the present study, we shall refer to these orthogonal basis functions in $\Phi$ as `controllability modes'. We note that if (\ref{general_dyn_sys_1}) can generate $\boldsymbol{u}$ whose covariance operator is identical to that from numerical simulations or experiments, the POD and controllability modes become identical.  

\subsection{Balanced truncation modes}\label{sec:2.3}
Instead of computing the most energetic structure in the output, balanced truncation produces a set of biorthgonal modes that are ranked in terms of dynamical importance \citep{mooreBPOD,zhoukemin,kWilcox/2.1570}. To this end, we first introduce the observability Gramian $\boldsymbol{W}_o^q$, which is obtained by solving the following equation:
\begin{equation}
    \boldsymbol{A}^\dag \boldsymbol{W}_o^q + \boldsymbol{W}_o^q \boldsymbol{A}+ \boldsymbol{C}^\dag \boldsymbol{C} = 0,
    \label{obsv_lyap}
\end{equation}
where $\boldsymbol{W}_o^q$ characterises the forcing structure $\boldsymbol{B}\boldsymbol{f}$ for the most energetic (or responsive) $\boldsymbol{u}$. In terms of forcing $\boldsymbol{f}$,
$\boldsymbol{W}_o = \boldsymbol{B}^\dag \boldsymbol{W}_o^q  \boldsymbol{B}$ is obtained instead. 

The controllability Gramian characterises the velocity structure of the most responsive $\boldsymbol{u}$, whereas the observability Gramian is for the structure of the forcing $\boldsymbol{f}$ associated with the most responsive $\boldsymbol{u}$. They typically appear in different forms, especially when the Reynolds number is high (see the discussion in \S\ref{sec:5.3}). To account for this difference, the balanced truncation introduces a coordinate transformation $\boldsymbol{T}$, defined by $\boldsymbol{u}=\boldsymbol{T}\boldsymbol{z}$, and subsequently seeks the $\boldsymbol{T}$ that makes $\boldsymbol{W}_c$ and $\boldsymbol{W}_o$ are identical (i.e. balancing): 
\begin{subeqnarray}
    \boldsymbol{W}_c &\mapsto& \boldsymbol{T}^{-1} \boldsymbol{W}_c (\boldsymbol{T}^{-1})^H,  \\ 
    \label{Wc_transform1}
    \boldsymbol{W}_o &\mapsto& \boldsymbol{T}^H\boldsymbol{W}_o \boldsymbol{T}.
    \label{Wo_transform1}
\end{subeqnarray}
It can be shown that such a coordinate exists, and the controllability and observability Gramian operators in the transformed coordinate are diagonal. In practice, the transformation $\boldsymbol{T}$ is obtained by computing the right eigenvector of $\boldsymbol{W}_c\boldsymbol{W}_o$: 
\begin{equation}
   \boldsymbol{W}_c\boldsymbol{W}_o\boldsymbol{T} = \boldsymbol{T}\Lambda, 
   \label{right_eigenvec}
\end{equation} 
where the transformation $\boldsymbol{T}$ provides a set of basis functions (i.e. $\Phi=\boldsymbol{T}$) known as the balanced truncation modes, and the diagonal entries in $\Lambda$ are the eigenvalues which provide their dynamical importance. 
Unlike the POD and controllability modes, the balanced truncation modes are not orthogonal to each other. Therefore, to apply a Galerkin projection with them, another set of functions that are orthogonal to the balanced truncation modes is required. This can be obtained by calculating the left eigenvector of $\boldsymbol{W}_c\boldsymbol{W}_o$, denoted by
$\boldsymbol{S}(\equiv \Psi=[ \psi_1(\boldsymbol{x})~\psi_2(\boldsymbol{x})~ ...])$. After a suitable normalisation of $\phi_i$ and $\psi_i$, the biorthogonality between $\Phi$ and $\Psi$ can be set to be satisfied: $\langle \phi_i, \psi_j \rangle = \delta_{ij}$.

\section{Reduced-order models for plane Couette flow}\label{sec:3}
As a benchmark case, we consider plane Couette flow of incompressible fluid with density $\rho$ and kinematic viscosity $\nu$. The two parallel walls move in a direction opposite to each other with a stream-wise velocity of $\pm U_w$, and are separated by a distance $2h$. All variables are dimensionless with $U_w$, $h$ and $\nu$, such that the resulting Reynolds number is defined to be $Re=U_wh/\nu$. The dimensionless streamwise, wall-normal and spanwise directions are denoted by $x$, $y$ and $z$, respectively.

To compute POD modes and assess the performance of the ROMs, a direct numerical simulation (DNS) is performed using an open source code, \texttt{channelflow} \citep{GibsonHalcrowCvitanovicJFM08}. The code employs the Fourier-Galerkin method for the discretisation in the streamwise and spanwise directions and the Chebyshev-tau method in the wall-normal direction. The computational domain is defined as $(x,y,z) \in [0, 1.75\pi)\times [-1,1]\times [0, 1.2\pi)$ and is close to the minimal flow unit of \cite{Hamilton_Kim_Waleffe_1995}. The resolution of the grid is $(N_x, N_y, N_z) = (32, 65, 32)$ after dealiasing. The Reynolds number considered is $Re=500$, and the corresponding friction Reynolds number is obtained as $Re_\tau (\equiv u_\tau h/\nu) \approx 34$ ($u_\tau$ is the friction velocity). 


\subsection{POD modes}
The POD modes in plane Couette flow are computed following the procedure described in \cite{Smith2005} and \cite{Chua_Khoo_Chan_Hwang_2022}. Since the flow is translationally invariant in the streamwise and spanwise directions, the POD modes are given in the form of the Fourier modes in those directions:
\begin{equation}\label{eq:3.1}
\phi_i(x,y,z)=\frac{1}{\sqrt{L_x L_z}}\hat{\phi}_{n_x,n_z}^{(n)}(y)\textrm{exp}(\textrm{i}( n_x \alpha_0 x+n_z\beta_0 )),
\end{equation}
where $n=1,2,...,N_y$, $\alpha_0=2\pi/L_x$, $\beta_0=2\pi/L_z$, and ($n_x$,$n_z$) is a pair of integers indicating each Fourier mode. Here, the index $i$ for the $i$-th POD mode on the left-hand side can now be suitably defined in terms of $(n_x,n_z,n)$. The covariance operator in (\ref{POD_eig}) can then be formulated in terms of individual Fourier modes, significantly reducing its size. This covariance operator is also computed by imposing the following discrete symmetries in plane Couette flow: 
\begin{subeqnarray}
    \mathcal{I}(u,v,w,p)[x,y,z,p] &=& (u,v,w,p)[x,y,z,p], \\
    \mathcal{P}(u,v,w,p)[x,y,z,p] &=& (-u,-v,-w,p)[-x,-y,-z,p], \\
    \mathcal{R}(u,v,w,p)[x,y,z,p] &=& (u,v,-w,p)[x,y,-z,p], \\
    \mathcal{RP}(u,v,w,p)[x,y,z,p] &=& (-u,-v,w,p)[-x,-y,z,p],
\end{subeqnarray}
where $p$ is pressure. Here, the group $\mathcal{I}$ indicates the identity transformation, $\mathcal{P}$ is a point reflection about the origin, $\mathcal{R}$ is the rotation about the plane of $z=0$ and $\mathcal{RP}$ is a 180$^\circ$ rotation about the $z$-axis. For further details, the reader may refer to \cite{Smith2005} and \cite{Chua_Khoo_Chan_Hwang_2022}.

\subsection{Linear operators}\label{sec:3.3}

For the computation of the controllability and balanced modes, here we define the linear operators introduced in \S\ref{sec:2.2}. We first decompose the flow field into a base flow $\boldsymbol{U} = (U(y),0,0)$ and perturbation/fluctuation $\boldsymbol{u} = (u,v,w)$. The dimensionless momentum equations for $\boldsymbol{u}$ are written as 
\begin{equation}\label{eq:3.3}
   \frac{\partial \boldsymbol{u} }{\partial t} + (\boldsymbol{U} \bcdot \nabla) \boldsymbol{u} + (\boldsymbol{u}\bcdot \nabla) \boldsymbol{U}   = 
    -\nabla p + \frac{1}{Re}\nabla^2 \boldsymbol{u}+\boldsymbol{N},
\end{equation}
where $\boldsymbol{N}$ is the nonlinear term given depending on the definition of $\boldsymbol{U}$.
Here, the ensemble average $\overline{(\cdot)}$ is equivalent to an average in time and in the streamwise and spanwise directions, as the given flow is statistically stationary and homogeneous in these wall-parallel directions. 

To model (\ref{eq:3.3}) in terms of a linear dynamical system, we consider various combinations of different base flows $\boldsymbol{U}$ and models for the nonlinear term $\boldsymbol{N}$, such that
\begin{subequations}\label{eq:3.4}
\begin{equation}\label{eq:3.4a}
U(y)=\left\{
\begin{array}{ll}
U_0(y) \\
\overline{U}(y)\\
\end{array}\right. \quad \textrm{and} \quad \boldsymbol{N}=\left\{
\begin{array}{ll}
\boldsymbol{f} \\
\nabla \cdot (\nu_t(\nabla \boldsymbol{u}+\nabla \boldsymbol{u}^T))+\boldsymbol{f},\\
\end{array}\right.
\end{equation}
where $U_0(y)(=y)$ is the laminar base flow, $\overline{U}$ the turbulent mean velocity obtained from DNS and $\boldsymbol{f} = [f_u,f_v,f_w]$ the forcing term providing a white noise in time. Here, the dimensionless eddy viscosity $\nu_t$ in $\boldsymbol{N}$ is given by a simple mixing length model \cite[]{Reynolds_Hussain_1972,HWANG_COSSU_2010_amplification}:
\begin{equation}
   -\overline{u v} = \nu_t \frac{d\overline{{U}}(y)}{dy}.
   \label{eddyViscHypo}
\end{equation}
\end{subequations}


Given the homogeneity in the streamwise and spanwise directions, we consider a Fourier mode representation of the velocity fluctuation/perturbation and forcing: \(\hat{\boldsymbol{u}}(y, t; \alpha, \beta) \text{e}^{\text{i}(\alpha x + \beta z)}\) and \(\hat{\boldsymbol{f}}(y, t; \alpha, \beta) \text{e}^{\text{i}(\alpha x + \beta z)} \), where \( \alpha = n_x\alpha_0\) and \(\beta= n_z\beta_0 \). 
Substituting these Fourier modes into (\ref{eq:3.3}) and eliminating the pressure fluctuation $p$ with the standard method, we obtain the following Orr-Sommerfeld-Squire system in terms of the Fourier modes of the wall-normal velocity and vorticity, $\hat{v}$ and $\hat{\eta}$:
\begin{subequations}\label{main_oss}
\begin{equation}
    \frac{\partial}{\partial t}\underbrace{\begin{bmatrix}
        \hat{v}\\
        \hat{\eta}
    \end{bmatrix}}_{\hat{\boldsymbol{q}}}
    =
    \underbrace{
    \begin{bmatrix}
        \Delta^{-1}\mathcal{L}_{OS} & 0 \\
        -i\beta U^\prime & \mathcal{L}_{SQ}
    \end{bmatrix}}_{\boldsymbol{A}}
    \begin{bmatrix}
        \hat{v} \\
        \hat{\eta}
    \end{bmatrix}
    +
    \underbrace{
    \begin{bmatrix}
        -i\alpha \Delta^{-1} \mathcal{D} & -k^2\Delta ^{-1} & -i\beta \Delta^{-1}\mathcal{D} \\
        i\beta & 0 & -i\alpha \\
    \end{bmatrix}
    }_{\boldsymbol{B}}
    \underbrace{\begin{bmatrix}
        \hat{f_u} \\
        \hat{f_v} \\
        \hat{f_w} \\
    \end{bmatrix}}_{\hat{\boldsymbol{f}}},
\end{equation}
where
\begin{eqnarray}
    \mathcal{L}_{OS} &=& -i\alpha(U\Delta - \mathcal{D}^2 U) + \nu_T \Delta^2 + 2\mathcal{D}\nu_t\Delta \mathcal{D} + \mathcal{D}^2\nu_t(\mathcal{D}^2 + k^2), \\
    \mathcal{L}_{SQ} &=& -i\alpha U + \nu_T \Delta + \mathcal{D}\nu_t \mathcal{D}
\end{eqnarray}
\end{subequations}
with the boundary conditions, $\hat{v}(y=\pm 1) =\mathcal{D}\hat{v}(y = \pm 1) = 0$ and $\hat{\eta}(y=\pm 1) = 0$. Here, $\mathcal{D}\equiv \partial \slash \partial y$, $\Delta = \mathcal{D}^2 - k^2$ and $k^2 = \alpha^2 + \beta^2$. Also, $\nu_T$ represents the total dimensionless eddy viscosity defined as $\nu_T=1/Re+\nu_t$, and the terms involving $\nu_t$ in (\ref{main_oss}) appear if $\boldsymbol{N}$ contains the related terms (see (\ref{eq:3.4a})). Using $\hat{\eta} = \partial \hat{{u}} \slash \partial z - \partial \hat{{w}}\slash{\partial x}$, the primitive variables are recovered with
\begin{equation}
    \underbrace{\begin{bmatrix}
        \hat{u} \\
        \hat{v} \\
        \hat{w} \\ 
    \end{bmatrix}}_{\hat{\boldsymbol{u}}}
    =
    \underbrace{\frac{1}{k^2}
    \begin{bmatrix}
        i\alpha & \mathcal{D}\\
        k^2 & 0 \\
        i\beta \mathcal{D} & i\alpha \\ 
    \end{bmatrix}
    }_{\boldsymbol{C}}
    \begin{bmatrix}
        \hat{v} \\
        \hat{\eta}
    \end{bmatrix}
    \label{oss_primitive_variable}.
\end{equation}
By replacing $\boldsymbol{u}$ and $\boldsymbol{q}$ in (\ref{general_dyn_sys_1}) with $\hat{\boldsymbol{u}}$ and $\hat{\boldsymbol{q}}$, it is now possible to generate the controllability and balanced modes in the form of (\ref{eq:3.1}). 

The eddy viscosity $\nu_t(y)$ is obtained using (\ref{eddyViscHypo}), but the mean velocity profile there was obtained from DNS with a slightly different domain size given by $(x,y,z) \in [0, 2\pi)\times [-1,1]\times [0, \pi)$. This is to avoid some unphysical behaviour in the mean velocity profile of the minimal flow unit, where $dU/dy<0$ was found around $y=0$ due to the highly restricted spatial domain. For the computation of the controllability and balanced modes, the wall-normal direction of (\ref{main_oss}) is discretised using a Chebyshev collocation method described in \cite{Weideman_Reddy} with $N_y=65$. The Lyapunov equations in (\ref{control_lyap}) and (\ref{obsv_lyap}) are solved using the \texttt{lyap} function in MATLAB. Lastly, for $\alpha = \beta = 0$, which corresponds to the flow averaged in the streamwise and spanwise directions, the linear operator $\boldsymbol{A}$ in (\ref{main_oss}) is singular. In this case, we instead used a set of the Stokes modes, the orthogonal basis functions obtained from the viscous diffusion operator, as in \cite{PhysRevFluids.7.L102601} and \cite{McCormack_Cavalieri_Hwang_2024}.

\begin{table}
  \begin{center}
\def~{\hphantom{0}}
  \begin{tabular}{lccc}
        Case~ &  $\phi_i$  &   $U$ & $\nu_t$  \\[5pt]
        
       C-LNL~   & ~Controllability modes~  & ~$U_0(y)$~ & ~0\\
       C-LNT~  & ~Controllability modes~ & ~$\overline{{U}}(y)$~ & ~0\\
       C-LNTe~  & ~Controllability modes~ & ~$\overline{{U}}(y)$~ & ~(\ref{eddyViscHypo})\\[3pt]
       
       BT-LNL~ & ~Balanced truncation modes~ & ~$U_0(y)$~ & ~0 \\
       BT-LNT~ & ~Balanced truncation modes~ & ~$\overline{{U}}(y)$~ &  ~0 \\
       BT-LNTe~ &~ Balanced truncation modes~ & ~$\overline{{U}}(y)$~ & ~(\ref{eddyViscHypo})\\ [3pt]
       
       POD~   & ~POD modes of turbulent state~ & ~ Not applicable ~& ~Not applicable \\
       
  \end{tabular}
  \caption{List of basis functions used for ROMs in the present study.}
  \label{model_summary}
  \end{center}
\end{table}

\subsection{Galerkin projection}
Using each set of POD modes from the turbulent state and controllability/balanced modes from various forms of (\ref{main_oss}) depending on the base flow $U$ and the nonlinear term model $\boldsymbol{N}$, we apply a Galerkin projection following the procedure described in \cite{PhysRevFluids.7.L102601} based on the Navier-Stokes equations for flow fluctuations around the laminar solution $U_0$. The list of basis functions and their construction approaches are summrised in table \ref{model_summary}. Notice that although different base flows were considered to obtain the basis functions, the Galerkin projection is constructed here based on fluctuations around the laminar solution $U_0$; thus, the Galerkin procedure is the same for all cases studied in table \ref{model_summary}, and the only change between ROMs is the choice of basis functions. For the Galerkin projection, the test functions used for the POD and controllability modes were themselves, resulting in
\begin{eqnarray} \label{general_ROM_sys}
    \underbrace{\frac{d{a}_i}{dt}}_{d\boldsymbol{a}/dt} && = \underbrace{ \frac{1}{Re} \sum_j \langle \nabla^2 \phi_j, \phi_i \rangle a_j- \sum_j \langle [(\phi_j \cdot \nabla) \boldsymbol{U_0} + (\boldsymbol{U_0} \cdot \nabla)\phi_j] , \phi_i \rangle a_j}_{\boldsymbol{La}}\nonumber  \\
    &&- \underbrace{  \sum_j \sum_k \langle (\phi_j \cdot \nabla )\phi_k,\phi_i \rangle a_j a_k}_{\boldsymbol{N}(\boldsymbol{a})},
\end{eqnarray}
where $\boldsymbol{a}=[a_1~a_2~a_3,....., ]^T$ and $\boldsymbol{U}_0=(U_0(y),0,0)$. Here, similarly to (\ref{eq:3.1}), the index $i$ of $a_i$ can be expressed in terms of $(n_x,n_z,n)$ for the $n$-the POD modes obtained from the Fourier component of $(n_x,n_z)$: i.e. $a_i=a_{n_x,n_z}^{(n)}$. Finally, for the ROM based on balanced truncation, the test functions were replaced with the left eigenvectors of $\boldsymbol{W}_c\boldsymbol{W}_o$, $\boldsymbol{S}$(or $\boldsymbol{\Psi}$), defined in \S\ref{sec:2.3}. For further details on the Galerkin projection, the reader refers to \cite{PhysRevFluids.7.L102601}. 

\begin{figure}
    \centering
    \includegraphics[width=\linewidth]{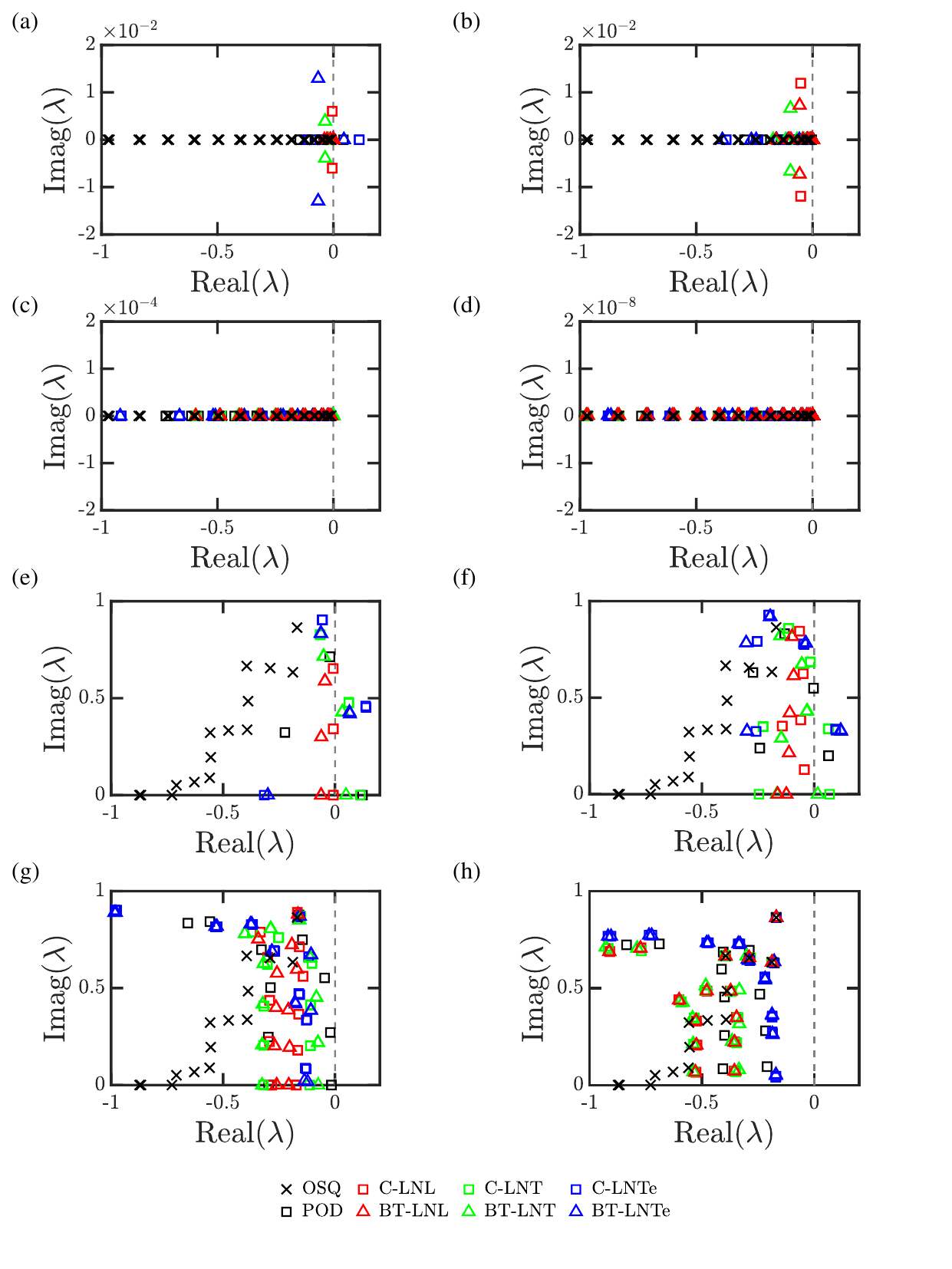}
    \caption{Eigenvalues of the linearised operator about the laminar base state: (a-d) $(n_x,n_z) = (0,1)$; (e-h) $(n_x,n_z) = (1,1)$. Here, (a,e) $N_y=5$; (b,f) $N_y=10$; (c,g) $N_y=20$; (d,h) $N_y=30$. Only top half of the complex plane is shown, as the eigenvalues are symmetric about $\textrm{Imag}(\lambda)=0$. In each case, dashed line indicates $\text{Real}(\lambda) = 0$. Here, the symbols denoted by OSQ represent the eigenvalues directly computed from the Orr-Sommerfeld-Squire system operator.}
    \label{eigen_11}
\end{figure}

\section{Results}\label{Result}

We present this section by dividing it into two subsections: one is on the convergence and performance of the ROMs around the laminar state, in which the linear stability and optimal transient growth computed with the ROMs are compared with those obtained with the full equations of motion, and the other is on their convergence and performance for turbulent state, where the turbulence statistics and the coherent structure dynamics, evaluated by temporal correlations, are examined against the corresponding DNS results.  

The computational domain of ROMs in the present study is the same as that of DNS described in \S\ref{sec:3}: $(x,y,z) \in [0, 1.75\pi)\times [-1,1]\times [0, 1.2\pi)$. Throughout the present study, we consider a single non-zero Fourier mode in the streamwise diection and two modes in the spanwise directions for the given computational domain: i.e. $n_x=0,\pm 1$ and $n_z=0, \pm 1, \pm 2$ in (\ref{eq:3.1}). The number of DoF of each of the ROMs is varied by changing the number of modes in the wall-normal direction, $N_y$, for each plane Fourier component. Here we consider $N_y = 5,10,20,30$. The corresponding total DoF are $N_{tot}=75,150,300,450$, respectively. 




\subsection{Laminar state}\label{sec:4.1}
We first examine the ROMs around the laminar base state, focussing on two pairs of wavenumbers, $(n_x,n_z) = (0,1)$ and $(n_x,n_z)= (1,1)$. These two pairs of wavenumbers were chosen due to their physical importance in the dynamics of the transition to turbulence \cite[]{Reddy1998}. The (0,1) mode provides the maximum (linear) transient growth associated with streaks, a key physical element of transition, and the excitation of the (1,1) mode leads to an oblique transition, which is energetically more efficient than the transition through the excitation of the (0,1) mode. We also note that the oblique transition scenario has recently been shown to be initiated by the Orr mechanism \cite[]{Jiao2021}, and is physically very similar to the transition triggered by the so-called `minimal seed \cite[i.e. the minimal energy initial condition for transition; for a review, see][]{Kerswell2018}. 

Figure \ref{eigen_11} shows the eigenvalues of the linearised operator for the two pairs of wavenumbers, on increasing the degree of freedom of the ROMs in the wall-normal direction, $N_y$. Specifically, a subset of linear operator $\boldsymbol{L}$ in (\ref{general_ROM_sys}) for each pair of the wavenumbers is extracted to study its linear stability. Here, we note that the laminar base state in plane Couette flow is linearly stable. In all cases, the eigenvalues from the ROMs gradually converge to those obtained directly from the linearised Navier-Stokes (or the Orr-Sommerfeld-Squire) operator, as $N_y$ increases, and in particular, the first couple of least stable eigenvalues show excellent convergence. In the cases of C-LNL and BT-LNL, the degrees to which the eigenvalues converge as $N_y$ increases are also seen to be consistent with those reported in \cite{ilak/1.2840197}: for example, compare figure \ref{eigen_11} with their figure 5.

Of all the cases tested, it was found that the cases C-LNL and BT-LNL, where the basis functions are obtained from the linearised operator with the laminar base flow and molecular viscosity, do not generate unstable eigenvalues for all examined $N_y$, indicating the preservation of the linear stability of the laminar base state even with very severe truncations. However, this was not necessarily the case for the ROMs with all the other basis functions, which are obtained with some information from turbulent flows. In this case, at $N_y \lesssim 20$, the laminar base state of the ROMs often produced linear instabilities, which are not desirable for modelling the state space around the laminar base flow: for example, $(n_x,n_z)=(0,1)$ with $N_y=5$ (figure \ref{eigen_11}a) and $(n_x,n_z)=(1,1)$ with $N_y=5,10$ (figures \ref{eigen_11}e and f). When increasing $N_y$ further, the first few eigenvalues are gradually captured well by the ROMs (figures \ref{eigen_11}d,f). Eventually, many leading eigenvalues of all ROMs converged to those of the linearised Navier-Stokes (or the Orr-Sommerfeld-Squire) operator with excellent accuracy when $N_y\gtrsim 100$ (not shown). 

\begin{table}
    \centering
    \begin{tabular}{l c c c c c c c}
    $(n_x,n_z)$~& ~POD~     & ~C-LNL~ & ~C-LNT~ & ~C-LNTe~ & ~BT-LNL~ & ~BT-LNT~ & ~BT-LNTe~ \\ [5pt] 
    (0,$\pm1$)~ & ~9 (1,6)~ & ~1~ & ~4~ & 6 & 1 & 5 & 9 \\
    (0,$\pm2$)~ & ~6 (1)~ & ~1~ & ~5~ & 8(6) & 1  & 9 (3) & 11 \\
    ($\pm 1$,0)~ & ~1~ & ~1~ & ~1~ & ~1~ & 1  & 6 (1) & 4 (1) \\
    (1,$\pm1$)~ & ~19 (1,3)~ & ~1~ & ~16 (14)~ & 17 (9,11,13,15) & 1 & 15 (13) & 15 (9,11) \\
    (1,$\pm2$)~  & ~18 (2)~  & ~1~ & ~14~ & ~19 (11,13,15,17)~ & 1 & 15 & 19 (1,2,15,17) \\
    (-1,$\pm1$) & ~19 (1,3)~ & ~1~ & ~16 (14)~ & ~17 (9,11,13,15)~ & 1 & 15 (13) & 15 (9,11) \\
    (-1,$\pm2$) & ~18 (2)~ & ~1~ & ~14~ & ~19 (11,13,15,17)~ & 1 & 15 & 19 (15,17) \\
    \end{tabular}
    \caption{The minimum $N_y$ greater than or equal to which laminar base state becomes stable. The number inside parenthesis indicates some $N_y$ at which the laminar state also shows stability.}\label{tab2}
\end{table}

Lastly, table \ref{tab2} summarises the minimum $N_y$ required to retain the linear stability of the laminar base state for all $(n_x,n_z)$ incorporated in the present ROMs. In all wavenumber pairs examined, cases C-LNL and BT-LNL consistently reproduced the linear stability of the laminar base state only with $N_y=1$. However, it was found that other ROMs require a substantially large value of $N_y$ for that.

\begin{figure}
    \centering
    \includegraphics[width=\linewidth]{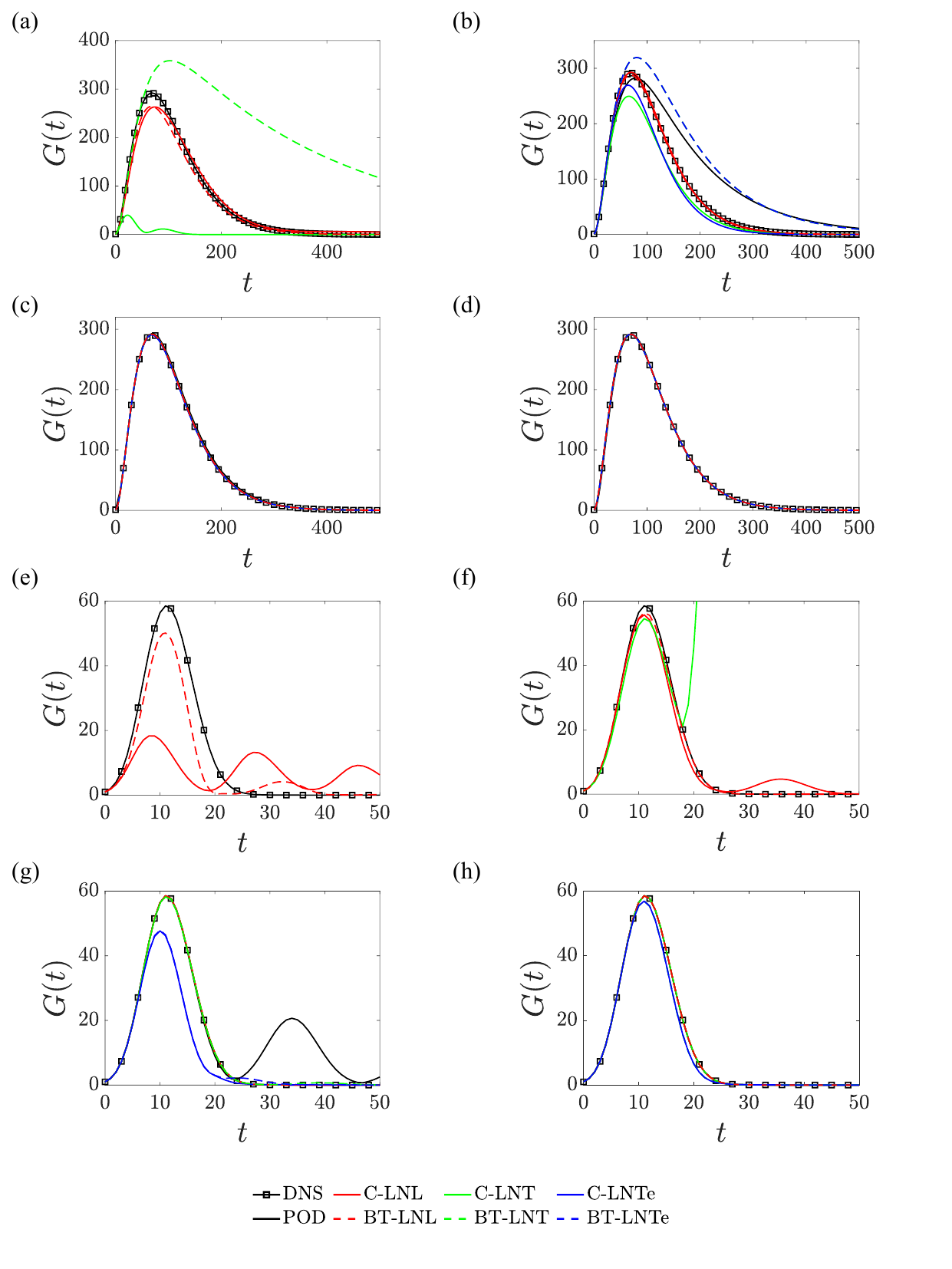}
    \caption{Optimal transient growth around laminar base flow for (a-d) $(n_x,n_z)=(0,1)$ and (e-h) $(n_x,n_z)=(1,1)$: (a,e) $N_y=5$; (b,f) $N_y=10$; (e,g) $N_y=20$; (d,h) $N_y=30$. Here, the lines missing indicate the cases that the corresponding ROM exhibit either rapid exponential growth (i.e. instability) or decay (not sufficient transient growth). In (c,d), the line closely overlaps with each other.}
    \label{rom_transient_01}
\end{figure} 

We now consider the optimal transient growth computed from the ROMs for each wavenumber \cite[e.g.][]{Schmid2002}. The optimal transient growth is defined as
\begin{equation}
    G(t{;}n_x,n_z) \equiv \lVert \boldsymbol{C}e^{\boldsymbol{L}t}\boldsymbol{B} \rVert^2=\max_{\boldsymbol{u}_0\not= 0} \frac{\lVert \boldsymbol{u} (t{;}n_x, n_z)  \rVert^2}{\lVert \boldsymbol{u}_0 (n_x, n_z)  \rVert^2},
    \label{rom_transient_eqn}
\end{equation}
where $\boldsymbol{u}_0$ is the initial condition. We note that optimal transient growth is also the $L_2$-norm of the state transition operator evaluated at every $t$, as shown above. 

The optimal transient growth for the linearised Navier-Stokes operator and the ROMs was calculated using the standard method \cite[]{Schmid2002}: for the linearised operator, the eigenfunction expansion technique was used, and, for the ROMs, the projected linear operator $\boldsymbol{L}$ was used with their basis functions. The transient growth is reported in figure \ref{rom_transient_01} for the two wavenumber pairs considered: $(n_x,n_z) = (0,1)$ and $(n_x,n_z)= (1,1)$. It is evident that the best performing ROM is the case BT-LNL (the ROM using balanced truncation modes obtained with laminar base flow and molecular viscosity), consistent with the previous finding of \cite{ilak/1.2840197}. The C-LNL case also performs fairly well, as it only needs $N_y \approx 10$ to capture the transient growth reasonably well. Similarly to the eigenvalue assessment in figure \ref{eigen_11}, the ROMs, whose basis functions are constructed using the information from turbulent flows, do not perform well, and this is evident from the results for $N_y\lesssim 10$. In all cases, the ROMs reproduced the optimal transient growth very well when $N_y=30$, although the C-LNTe and BT-LNTe cases still showed a slight deviation for $(n_x,n_z)= (1,1)$ (figure \ref{rom_transient_01}h).

\subsection{Turbulent state}

\begin{figure}
    \centering
    \includegraphics[width=\linewidth]{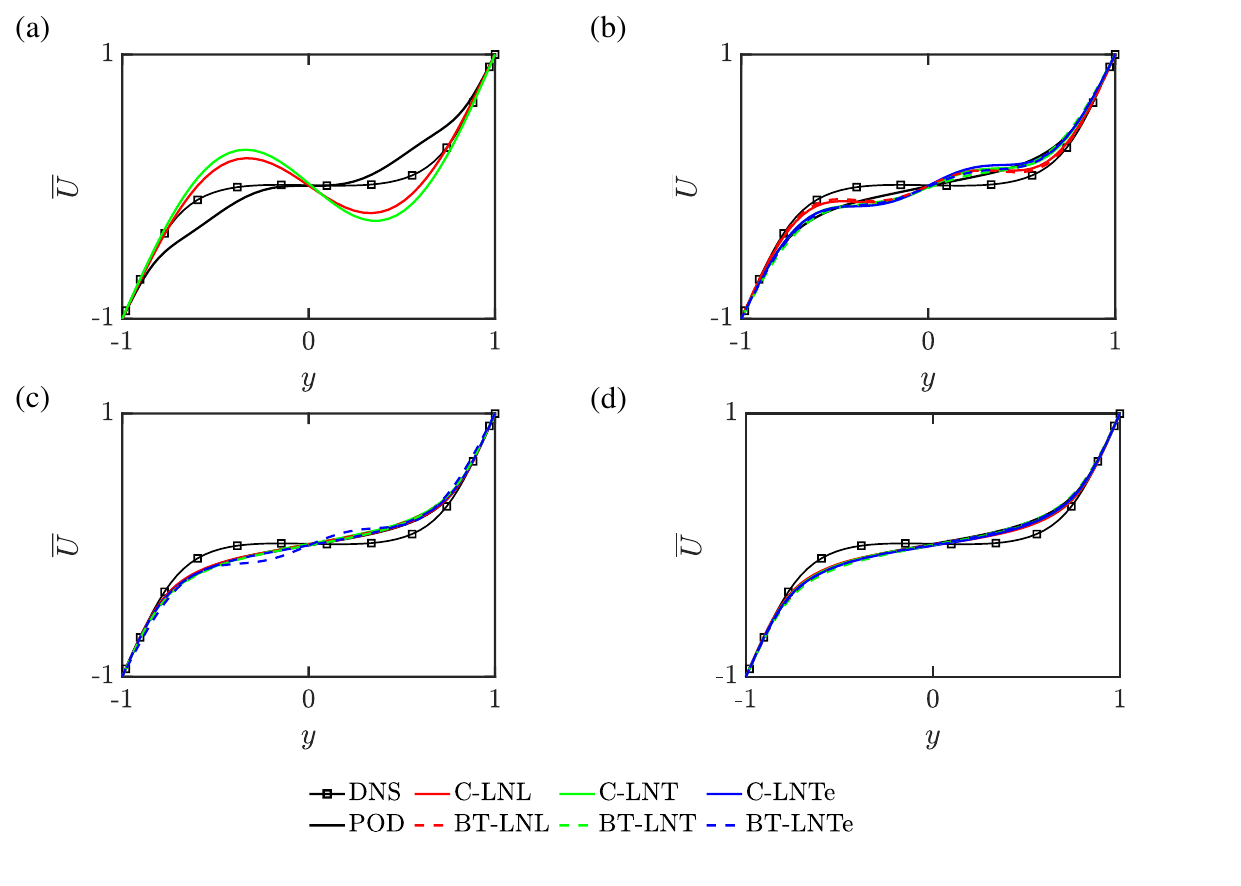}
    \caption{Mean streamwise velocity of turbulent state: (a) $N_y=5$; (b) $N_y=10$; (c) $N_y=20$; (d) $N_y=30$. In (a), the cases missing from the plot either show the solution rapidly blowing-up or relaminarisation (see texts). In (c,d), the ROM profiles are hardly discernible from each other due to their convergence.}
    \label{Umean_compiled}
\end{figure}
\begin{figure}
    \centering
    \includegraphics[width=\linewidth]{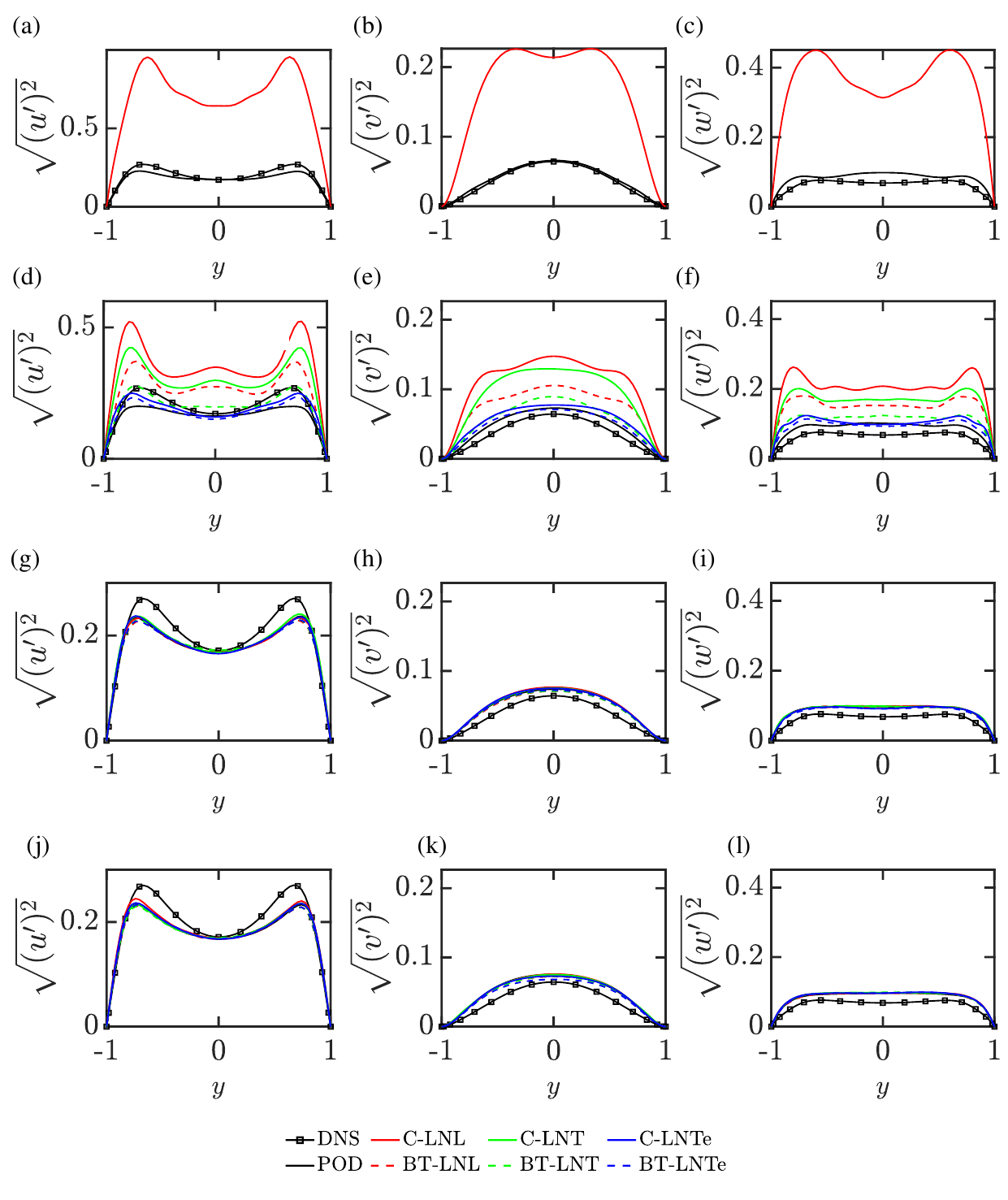}
    \caption{Root-mean-square (a,d,g,j) streamwise, (b,e,h,k) wall-normal and (c,f,i,l) spanwise velocity of turbulent state: (a-c) $N_y=5$; (d-f) $N_y=10$; (g-i) $N_y=20$; (j-l) $N_y=30$. In (a), the cases missing from the plot either show the solution rapidly blowing-up or relaminarisation (see texts), except the C-LNT case that is not shown due to huge deviations from the DNS profiles.}
    \label{rom_rms}
\end{figure}

We now examine the modelling capability of the ROMs for turbulent state. The ROMs were integrated in time using the function \texttt{ode45} in MATLAB. The initial condition for the ROMs in (\ref{general_ROM_sys}) is first generated as a random sequence with $a_i \in (0,0.1)$. Equation (\ref{general_ROM_sys}) is subsequently integrated to see if it reaches a statistically stationary state. 
When $N_y=5$, not all ROMs were able to generate a statistically stationary turbulent state. Only the cases C-LNL, C-LNT and POD can generate a statistically stationary turbulent state. The case BT-LNL, which was able to capture the stability of the laminar state the most effectively, showed rapid relaminarisation, and BT-LNT, C-LNTe, and BT-LNTe generated solutions that were rapidly blown up. Regardless of the stability of the laminar base state, when $N_y=10$, all ROMs were able to generate a stationary turbulent state. Finally, when $N_y\gtrsim 20$, all ROMs were able to retrieve the linearly stable laminar state (table \ref{tab2}) and a stationary turbulent state. 

\subsubsection{Turbulence statistics}
Figure \ref{Umean_compiled} reports the mean velocity profiles obtained by averaging over a period of time $T_{avg}=10000$. 
At the heaviest truncation ($N_y=5$; figure \ref{Umean_compiled}a), the mean velocity profiles from the C-LNL, C-LNT and POD cases, which were able to generate a statistically stationary turbulent state, are considerably different from the one obtained from DNS. In particular, C-LNL and C-LNT show negative values of $dU/dy$ and the related overshoot in the mean velocity profile $U(y)$ around $y=0$. When $N_y$ is doubled ($N_y=10$; figure \ref{Umean_compiled}b), all the ROMs generated a turbulent state, the mean of which appears to be very difficult to distinguish from each other. In this case, the wall-normal locations, at which $dU/dy<0$, are very small or do not exist. Therefore, there is no visible overshoot in $U(y)$. Finally, for $N_y\gtrsim 20$, the mean velocity profiles of all the ROMs appear to converge into a single one, which shows only a small difference from the mean velocity profile obtained with DNS. This difference originates from the modelling setting of the ROMs, where we only considered a limited number of Fourier modes in the wall-parallel directions. In fact, when the number of Fourier modes is further increased, the mean velocity profile of ROMs converges to that of DNS, as shown by \cite{PhysRevFluids.7.L102601} for the C-LNL case with the basis functions obtained at $Re=100$. 



Root-mean-square (rms) velocity profiles are also shown in figure \ref{rom_rms}. For $N_y=5$ (figures \ref{rom_rms}a-c), the case POD shows remarkably good performance even with such a severe truncation. This is particularly true, as the profiles from the cases C-LNL and C-LNT significantly deviate from those obtained with DNS (the case C-LNT is not shown in figure \ref{rom_rms}a due to its huge deviation). When $N_y=10$ (figures \ref{rom_rms}d-f), the rms profiles of all ROMs begin to predict reasonably well compared to those of DNS. The POD case still predicts the rms profiles well, and the ROMs constructed using the mean velocity and the corresponding eddy viscosity (i.e. the cases C-LNTe and BT-LNTe) also almost equally perform well. All the other ROMs show a non-negligible difference in the rms profiles from those obtained with DNS, except the BT-LNT case, the prediction of which is only slightly worse than the POD, C-LNTe and BT-LNTe cases. 
Finally, at $N_y\gtrsim20$ (figures \ref{rom_rms}g-l), the velocity fluctuations of all ROMs converge to a profile that deviates slightly from the DNS statistics. Incorporating additional streamwise and spanwise Fourier harmonics in (\ref{eq:3.1}) improves the convergence of rms velocity profiles, as already discussed for mean velocity. 

Overall, the observations here highlight that the ROMs using the basis functions from the linearised Navier-Stokes operator with turbulent mean velocity and the eddy viscosity model (C-LNTe and BT-LNTe) generally perform well in predicting turbulence statistics. Their performance is comparable to that of the ROM with POD modes and superior to those using the modes obtained from the linearised Navier-Stokes operator without the eddy viscosity model. This suggests that adding the eddy viscosity model to the linearised Navier-Stokes operator would be useful for the generation of basis functions, especially at high Reynolds numbers, as they would potentially be able to keep the dimension of Galerkin-projection-based ROMs relatively low. 

\begin{figure}
    \centering
    \includegraphics[width= \linewidth]{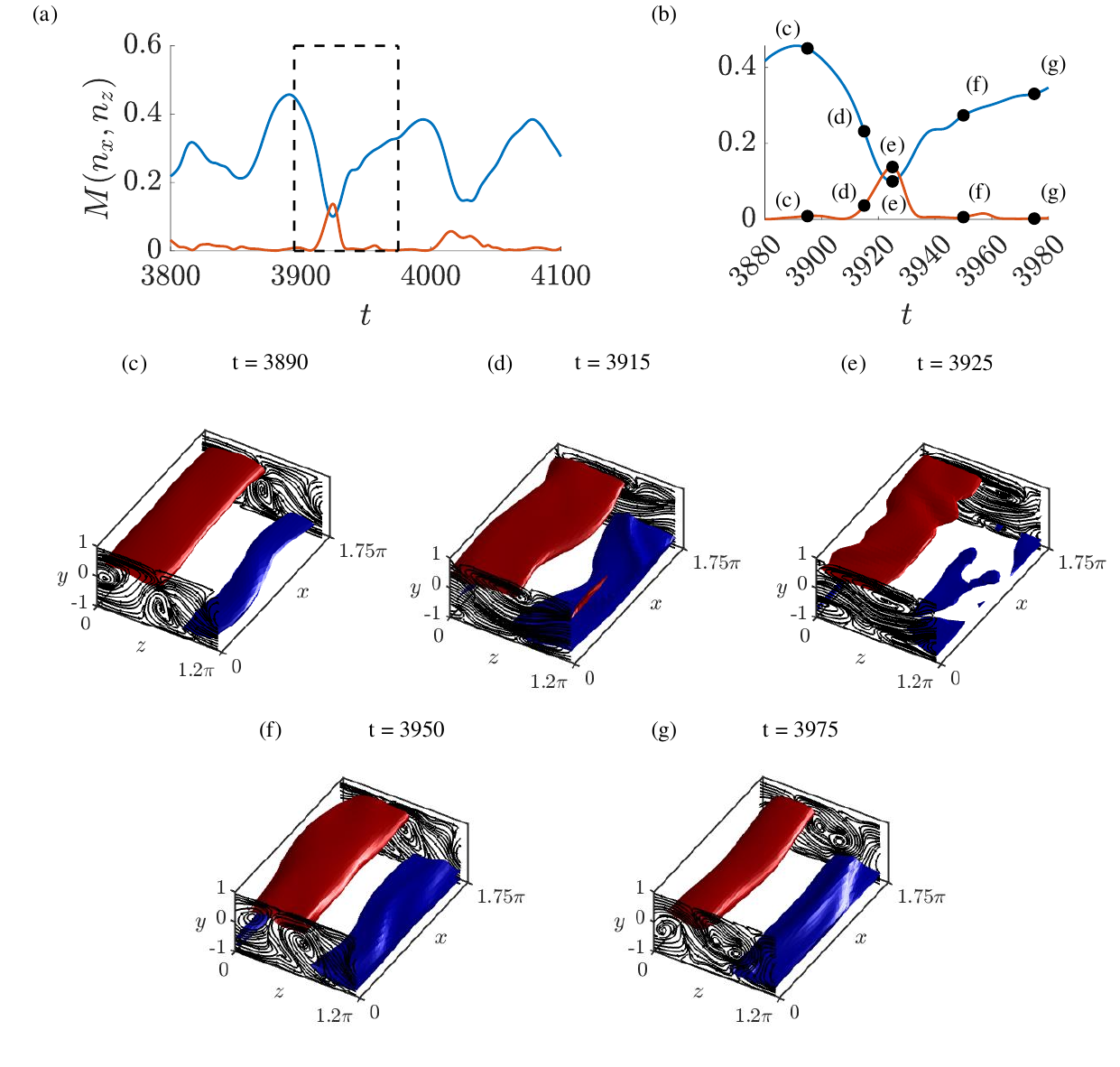}
    \caption{(a) Time trajectories of $M(0,1)$ and $M(1,1)$ and (b) their magnified view for $t \in [3880,3980]$ in DNS: blue, $M(0,1)$; orange, $M(1,1)$. (b)-(d) Flow field snapshots at different stages of SSP, with blue/red isosurfaces indicating $u = \pm 0.65$, respectively. Here, the streamlines on $x = 0$ and $x=1.75\pi$ indicate the spanwise and wall-normal velocity fields.}
    \label{dns_breakdown}
\end{figure}

\begin{figure}
    \centering
    \includegraphics[width=\linewidth]{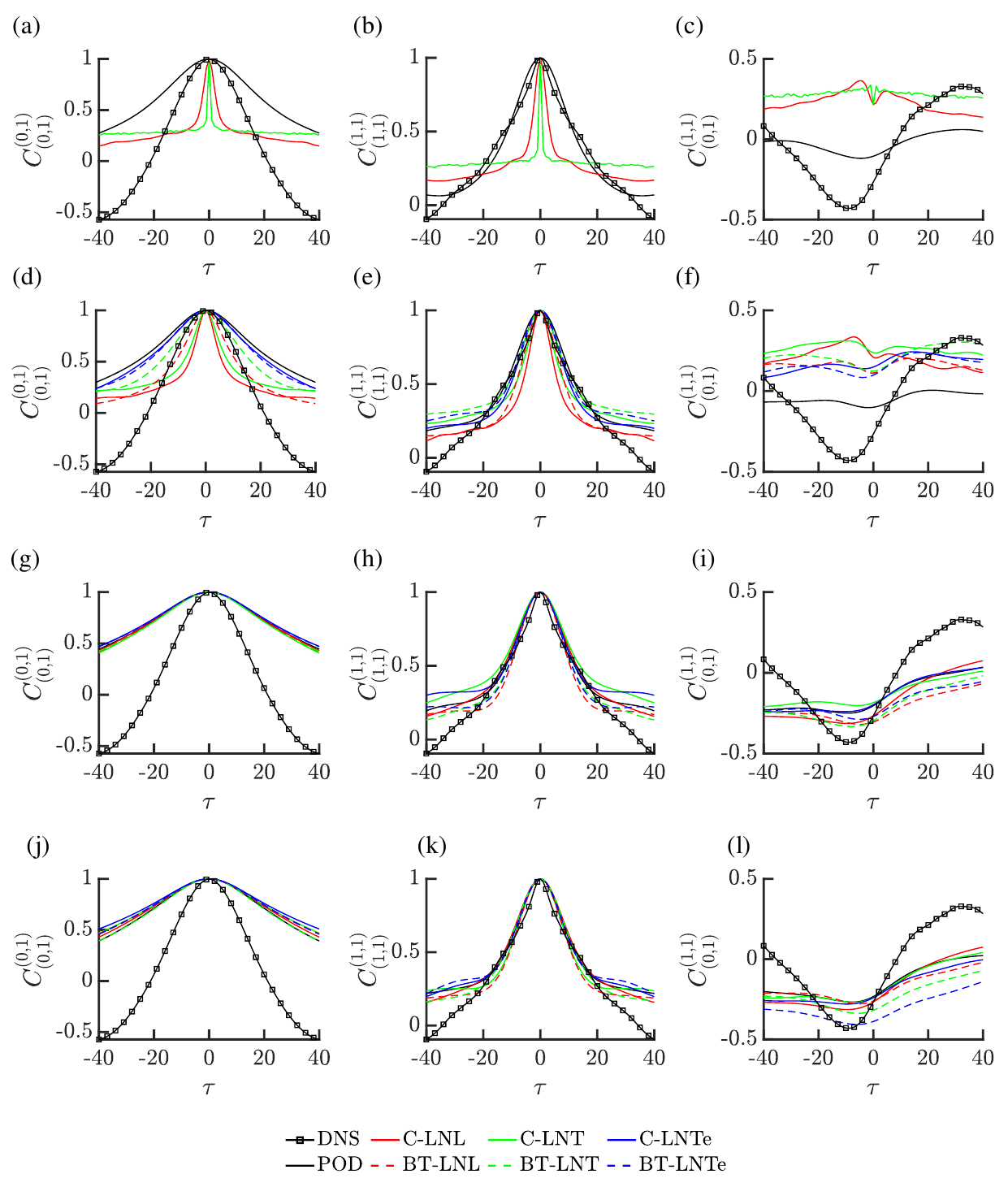}
    \caption{Auto- and cross-correlations of $M(0,1)$ and $M(1,1)$ for different ROMs: (a-c) $N_y=5$; (b-f) $N_y=10$; (g-i) $N_y=20$; (j-l) $N_y=30$.}
    \label{fig:xcor}
\end{figure}

\subsubsection{Dynamics of coherent structures}
Coherent structure dynamics in the minimal flow units of plane Couette flow has been well understood, since the pioneering analysis by \cite{Hamilton_Kim_Waleffe_1995}. Figure \ref{dns_breakdown} explains the coherent structure dynamics, often referred to as the `self-sustaining process'. There, the time evolution of the key observables is presented in figures \ref{dns_breakdown}(a,b), and the corresponding flow visualisations are shown in figures \ref{dns_breakdown}(c-g). The flow field is dominated by elongated streamwise velocity structures called streaks (figure \ref{dns_breakdown}c), the spanwise spacing of which is captured by $L_z$. The streaks are therefore well characterised by the energy of the plane Fourier mode, defined by
\begin{equation}
    M(t{;}n_x,n_z) = \int_{-1}^{1} \tilde{\boldsymbol{u}}^H(t,y;n_x,n_z)\tilde{\boldsymbol{u}}(t,y;n_x,y,n_z)~dy,
    \label{modal_energy}
\end{equation}
for $(n_x,n_z)=(0,1)$. The streaks gain their energy from mean shear and are amplified. When their energy is high enough (top one in figure \ref{dns_breakdown}c), they experience instability and/or transient growth, both typically involving sinuous meandering motion with the fundamental streamwise wavelength in the minimal flow unit (figure \ref{dns_breakdown}d). This meandering motion can be characterised by $M(t,n_x,n_z)$ with $(n_x,n_z)=(1,1)$. After experiencing the meandering motion, the streaks break down and reach a low-energy state, while generating streamise vortices (figure \ref{dns_breakdown}e). The streamwise vortices regenerate the streaks (figure \ref{dns_breakdown}f,g), and the physical processes described here repeat with the typical time scale of $T=80 \sim 100$ in the minimal flow unit of plane Couette flow.  To quantitative assess these temporal dynamics of the coherent structures in the ROMs, we introduce the temporal auto- and cross-correlations between $M(t;n_x,n_z)$:
\begin{equation}
    C_{(m_x,m_z)}^{(n_x,n_z)}(\tau) = \frac{\overline{ \Tilde{M}(t+\tau{;}n_x,n_z) \Tilde{M}(t{;}m_x,m_z)}}{\sqrt{\Tilde{M}^2(t{;}n_x,n_z)}\sqrt{\Tilde{M}^2(t{;}m_x,m_z)}},
\end{equation}
where $\Tilde{M}(t{;}n_x,n_z) = M(t{;}n_x,n_z) - \overline{M(t{;}n_x,n_z)}$ is the fluctuation around the mean. 

Figure \ref{fig:xcor} shows the correlations evaluated from the ROMs and subsequently compares them with those from DNS using $M(t;0,1)$ (streaks) and $M(t;1,1)$ (streak meandering motions). 
At the heaviest truncation ($N_y=5$; figures \ref{fig:xcor}a-c), among the three ROMs that exhibit no numerical instability (POD, C-LNL, C-LNT), the POD case was found to best capture the timescale of correlations obtained with DNS. In particular, $C_{(0,1)}^{(1,1)}$ was well reproduced with such a heavy truncation, and the negative values of $C_{(0,1)}^{(1,1)}$ for $\tau\lesssim 0$, which represent the phase difference between $M(0,1)$ and $M(1,1)$, were also observed. In contrast, the other two cases show auto-correlations with very short time scales compared to DNS, and their cross-correlations appear to be largely unphysical (e.g., the behaviour of correlation tails around $\tau \simeq \pm 40$). 


As the degree of freedom increases to $N_y=10$ (figures \ref{fig:xcor}d-f), the auto-correlations of the ROMs appear to improve, in addition to the fact that all ROMs can now generate a turbulent solution. In particular, the cases POD, C-LNTe and BT-LNTe show relatively well-converged auto-correlations (figures \ref{fig:xcor}d,e) compared to those for $N_y=30$ (figures \ref{fig:xcor}j,k). However, the negative values of $C_{(0,1)}^{(1,1)}$ for $\tau\lesssim 0$ are still not seen for all ROMs, except the POD case, suggesting that the ROM based on POD modes performs best in terms of capturing the coherent structure dynamics of turbulent flow. 
Finally, when $N_y\gtrsim 20$, all the correlation functions of the ROMs appear to be reasonably well converged, and their $C_{(1,1)}^{(1,1)}$ now also shows negative correlations for $\tau\lesssim 0$, consistent with DNS (figures \ref{fig:xcor}g-l). 

\section{Discussions} \label{discussion}
Thus far, we have explored the performance of several ROMs constructed with various forms of basis functions: POD modes from turbulent flows, and controllability and balanced truncation modes of the linearised Navier-Stokes operator considering different base flows (laminar base flow or turbulent mean flow) and incorporation of an eddy viscosity model. 
It has become evident that the performance and convergence of these ROMs around both laminar and turbulent state depends on the state space, the information of which is used to generate the different sets of the basis functions. Around the laminar state, the cases C-LNL and BT-LNL, which use the controllability and balanced truncation modes of the Navier-Stokes operator linearised about the laminar base flow, do not exhibit any linear instabilities even with only a single mode in the wall-normal direction ($N_y=1$). In particular, the BT-LNL case rapidly captures the transient growth, the linear route of the transition to turbulence, as $N_y$ increases ($N_y \lesssim 10$), consistent with previous findings of \cite{ilak/1.2840197}. In contrast, all other ROMs require at least $N_y \gtrsim 20$ to achieve the linear stability and transient growth of the laminar base state (table \ref{tab2} and figure \ref{rom_transient_01}). 

The turbulent state has been found to be best modelled with the ROM based on POD modes, which have often been criticised due to the nature accounting for only the energy content without considering the dynamics. It is also worth noting that the cases C-LNTe and BT-LNTe model turbulence statistics fairly well at relatively low $N_y$ (e.g. $N_y=10$; figures \ref{Umean_compiled}, \ref{rom_rms} and \ref{fig:xcor}), indicating that the modes from the eddy-viscosity-enhanced linearised Navier-Stokes operator with turbulent mean flow have some non-negligible advantages over the other operator-based modes. An advantage of the cases C-LNTe and BT-LNTe over the POD-based ROMs is that they do not require prior simulation or experimental data, as they are derived directly from the linearised Navier-Stokes system, with some elements from the turbulent state via the use of mean flow and eddy viscosity. However, unlike the laminar state, the advantage of using the balanced truncation modes does not appear to be evident, given the comparable performance of the C-LNTe and BT-LNTe cases in modelling the turbulent state. The advantage of using the balanced truncation modes appears more evident in case BT-LNT compared to case C-LNT, but both of the cases, obtained with the linearised Navier-Stokes operator with turbulent mean flow, do not perform better than their counterparts with the eddy viscosity model (cases C-LNTe and BT-LNTe).

\begin{table}
  \begin{center}
\def~{\hphantom{0}}
  \begin{tabular}{lccccc}
         &  \multicolumn{2}{c}{Laminar state} & \multicolumn{3}{c}{Turbulent state} \\             
        ROMs        &     ~Stability~      &   ~Transient growth~    &    ~Mean~   & ~Fluctuations~  & Correlations \\[5pt]
        C-LNL~    & ~Good~ &    ~Good~       &  ~Poor~   &   ~Poor~  &   ~Good~\\
        C-LNT~    & ~Poor~   & ~Poor~ &  ~Poor~   &  ~Poor~ &   ~Good~\\ 
        C-LNTe~   & ~Poor~   & ~Poor~ &  ~Good~   &  ~Good~ &   ~Good~\\ [3pt]
          
        BT-LNL~   & ~Best~   &  ~Best~  & ~Poor~ &  ~Poor~  &   ~Poor~ \\ 
        BT-LNT~   & ~Poor~   &   ~Poor~  & ~Good~ &  ~Good~  &   ~Good~ \\ 
        BT-LNTe~  & ~Poor~   &   ~Poor~  & ~Good~ & ~Good~ &   ~Good~ \\ [3pt]

        POD      & ~Poor~ &     ~Poor~     & ~Best~     & ~Best~ &   ~Best~ \\
       
  \end{tabular}
  \caption{Performance of the ROMs depending on the state to be modelled (see text for a further explanation).}
  \label{pf_summary}
  \end{center}
\end{table}

\begin{figure}
    \centering
    \begin{subfigure}{0.45\textwidth}
         \includegraphics[width=\textwidth]{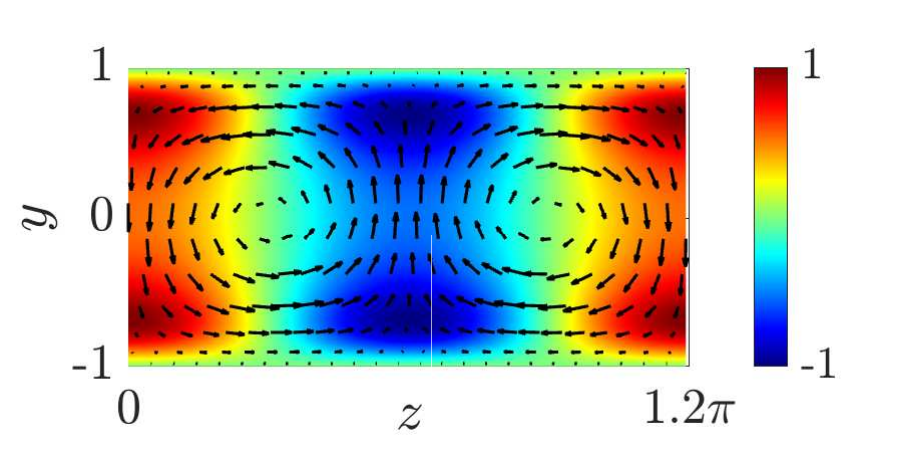}
         \caption{POD}
    \end{subfigure}
    \begin{subfigure}{0.45\textwidth}
         \includegraphics[width=\textwidth]{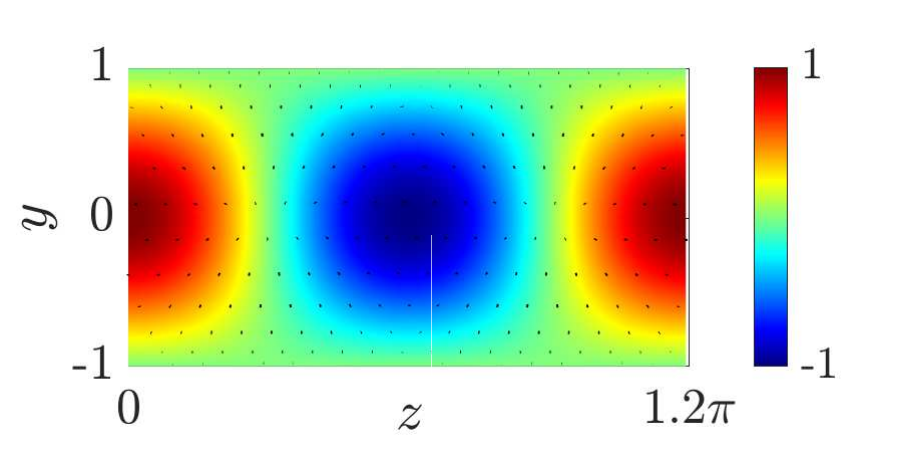}
         \caption{C-LNL}  
    \end{subfigure}
    \begin{subfigure}{0.45\textwidth}
         \includegraphics[width=\textwidth]{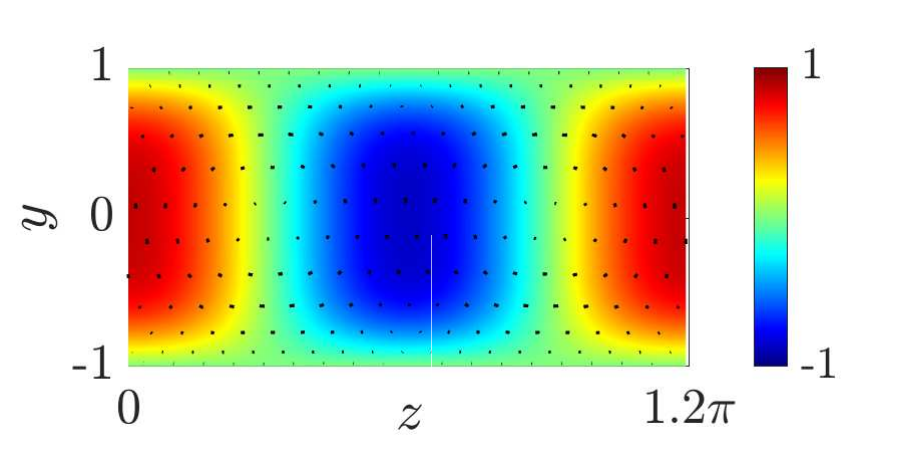}
         \caption{C-LNT}
    \end{subfigure}
    \begin{subfigure}{0.45\textwidth}
         \includegraphics[width=\textwidth]{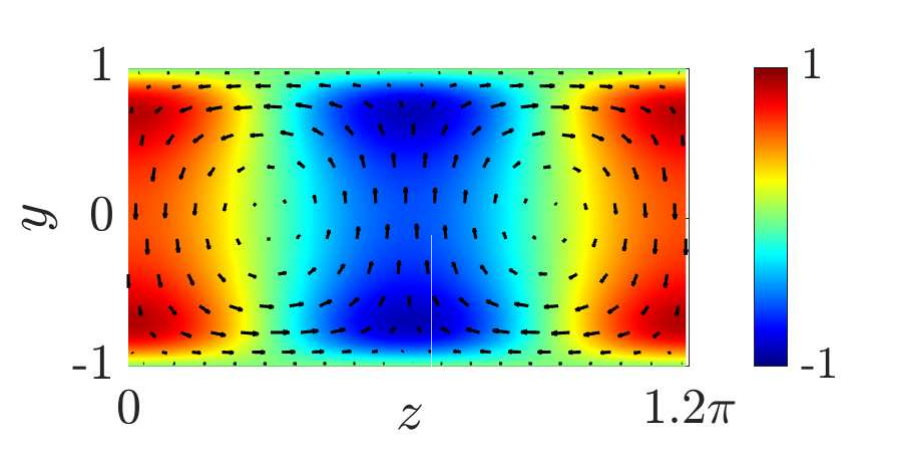}
         \caption{C-LNTe}
    \end{subfigure}
    \begin{subfigure}{0.45\textwidth}
         \includegraphics[width=\textwidth]{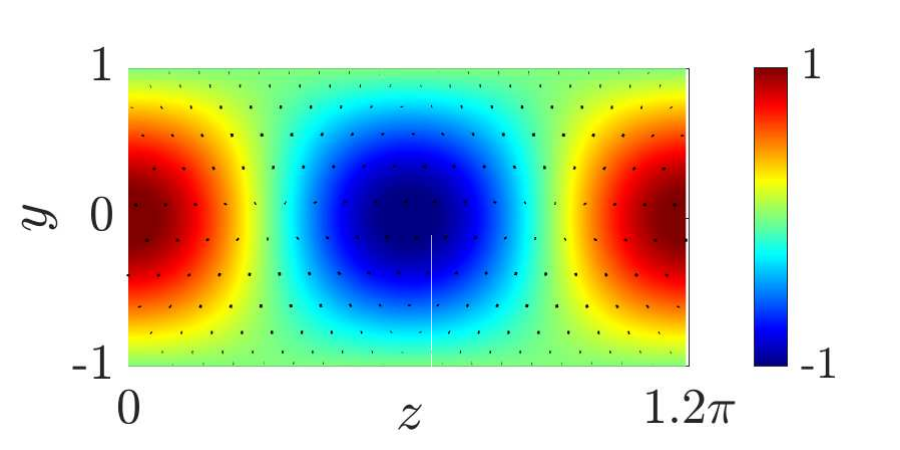}
         \caption{BT-LNL direct mode}
         
    \end{subfigure}
    \vspace{\fill}
    \begin{subfigure}{0.45\textwidth}
         \includegraphics[width=\textwidth]{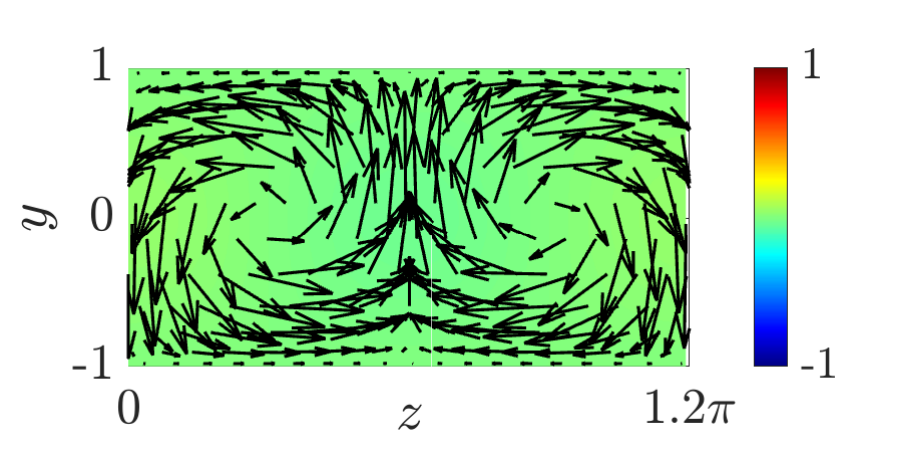}
         \caption{BT-LNL adjoint mode}
         
    \end{subfigure}
    \vspace{\fill}
    \begin{subfigure}{0.45\textwidth}
         \includegraphics[width=\textwidth]{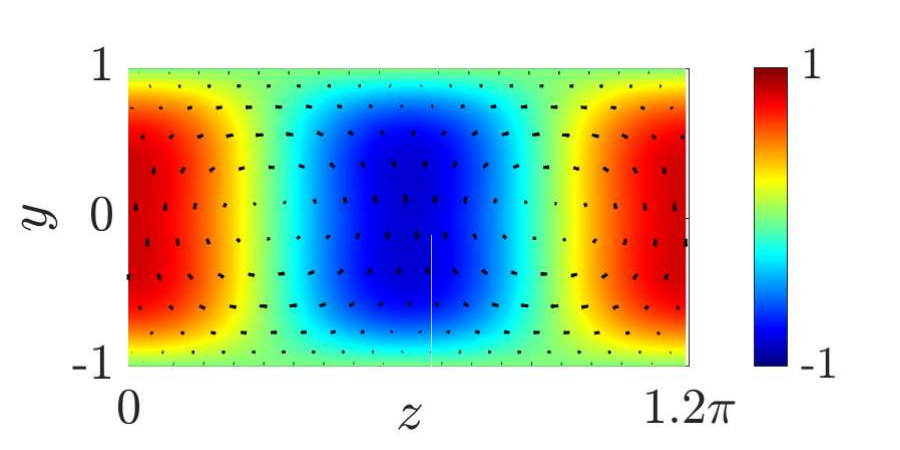}
         \caption{BT-LNT direct mode}

    \end{subfigure}
    \begin{subfigure}{0.45\textwidth}
         \includegraphics[width=\textwidth]{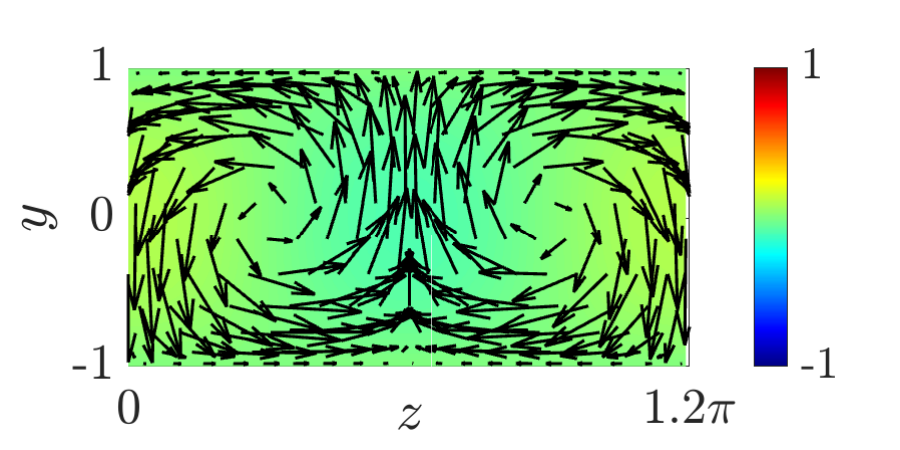}
         \caption{BT-LNT adjoint mode}
    \end{subfigure}
    \vspace{\fill}
    \begin{subfigure}{0.45\textwidth}
         \includegraphics[width=\textwidth]{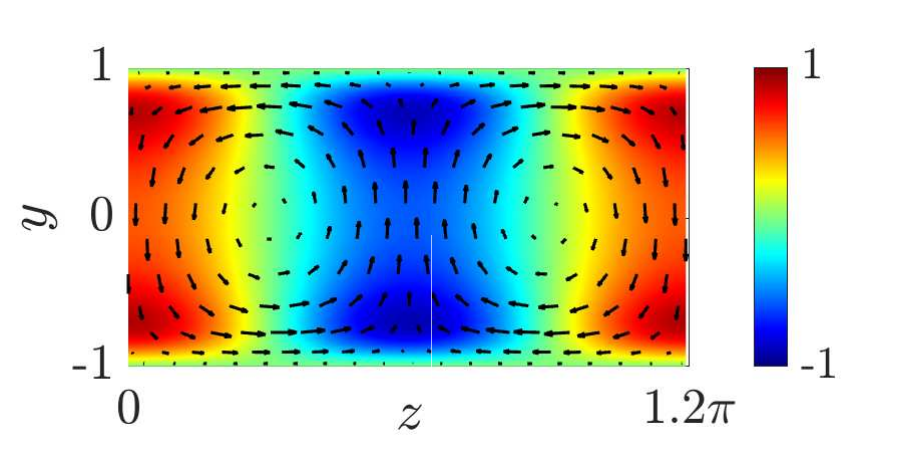}
         \caption{BT-LNTe direct mode}

    \end{subfigure}
    \begin{subfigure}{0.45\textwidth}
         \includegraphics[width=\textwidth]{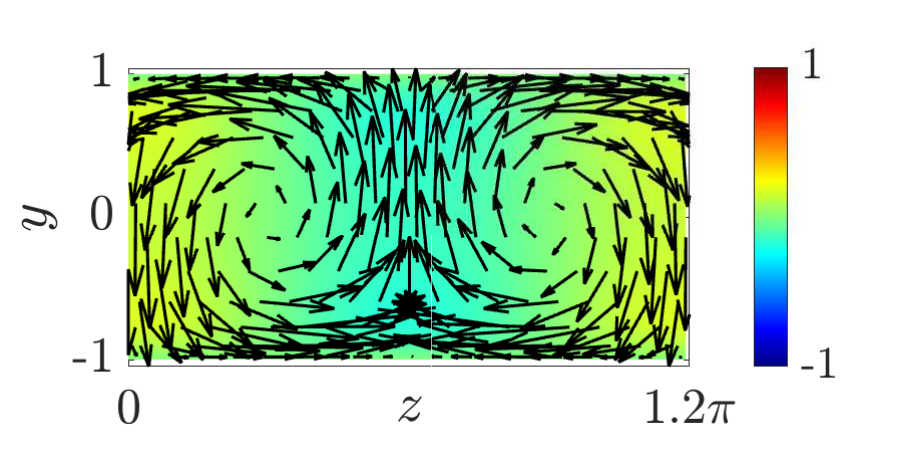}
         \caption{BT-LNTe adjoint mode}
    \end{subfigure}
    \caption{Visualisation of the leading mode of different basis functions for ($n_x,n_z) = (0,1)$ on the $y$-$z$ plane. The positive/negative contours indicate positive/negative streamwise velocity, and the vectors represents the wall-normal and spanwise velocity components. Each mode is normalised by its energy, and the size of the vectors are scaled to be the same for all the cases.}
    \label{basis_visual}
\end{figure}

\begin{figure}
    \centering
    \begin{subfigure}{0.45\textwidth}
         \includegraphics[width=\textwidth]{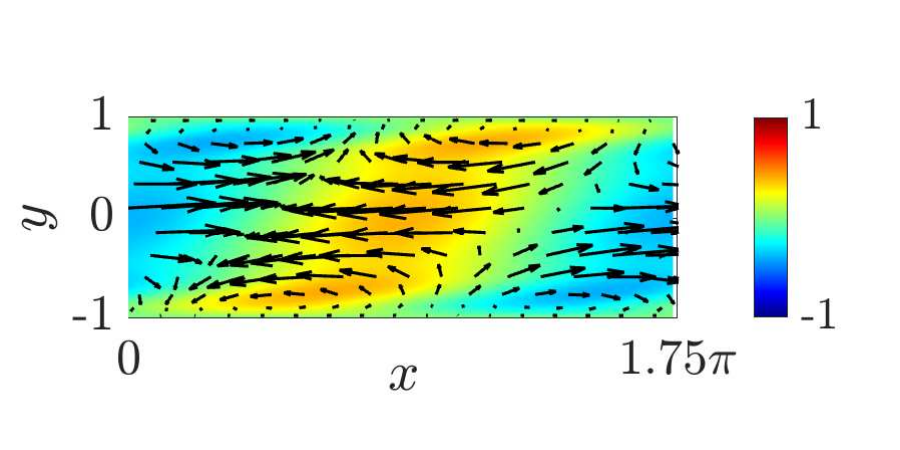}
         \caption{POD}
    \end{subfigure}
    \begin{subfigure}{0.45\textwidth}
         \includegraphics[width=\textwidth]{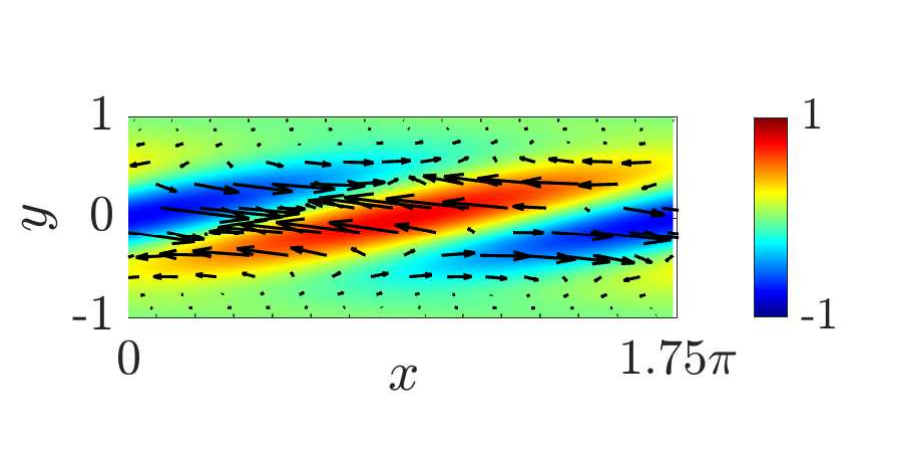}
         \caption{C-LNL}  
    \end{subfigure}
    \begin{subfigure}{0.45\textwidth}
         \includegraphics[width=\textwidth]{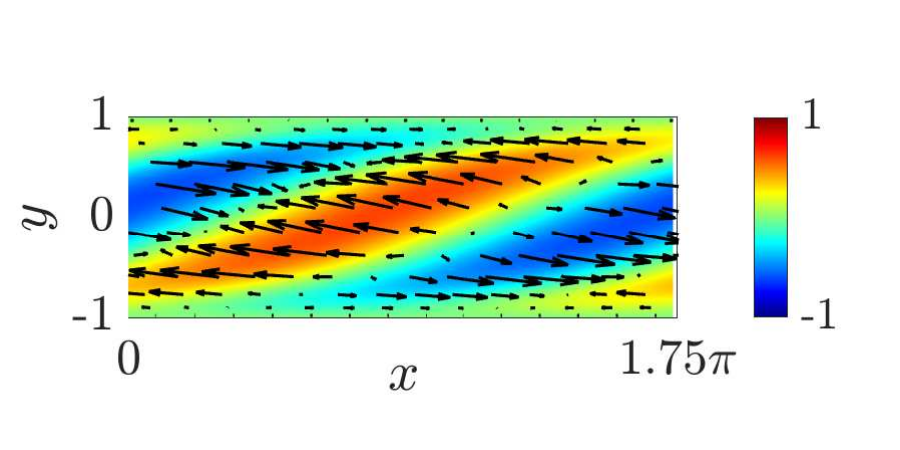}
         \caption{C-LNT}
    \end{subfigure}
    \begin{subfigure}{0.45\textwidth}
         \includegraphics[width=\textwidth]{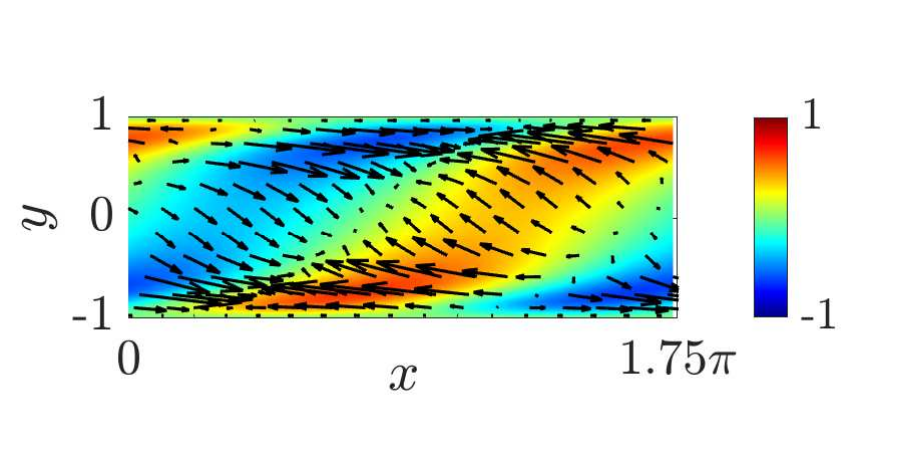}
         \caption{C-LNTe}
    \end{subfigure}

    \begin{subfigure}{0.45\textwidth}
         \includegraphics[width=\textwidth]{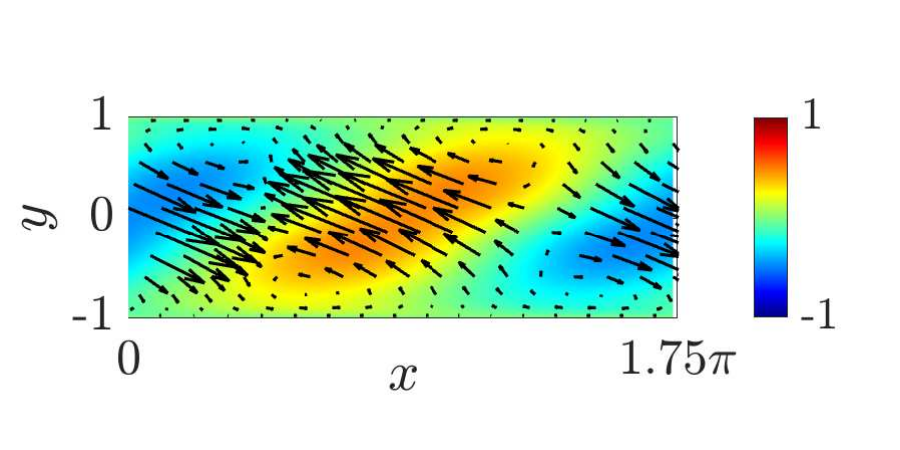}
         \caption{BT-LNL direct}
         
    \end{subfigure}
    \vspace{\fill}
    \begin{subfigure}{0.45\textwidth}
         \includegraphics[width=\textwidth]{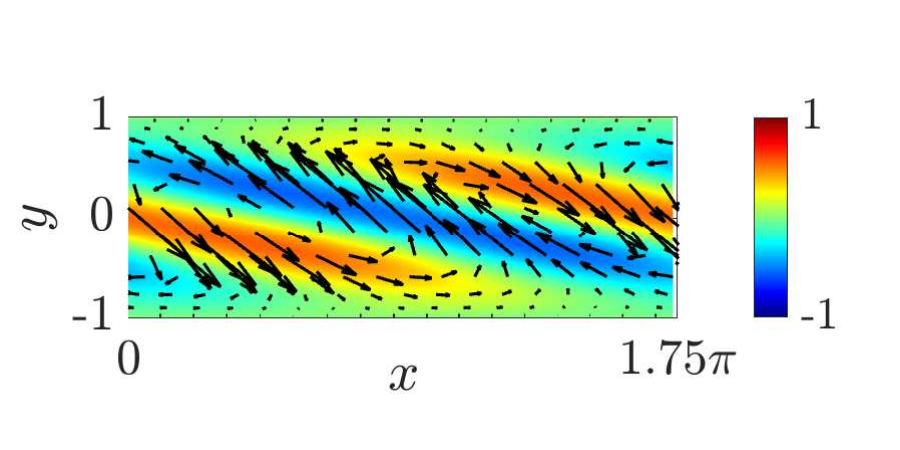}
         \caption{BT-LNL adjoint}
         
    \end{subfigure}
    \vspace{\fill}
    \begin{subfigure}{0.45\textwidth}
         \includegraphics[width=\textwidth]{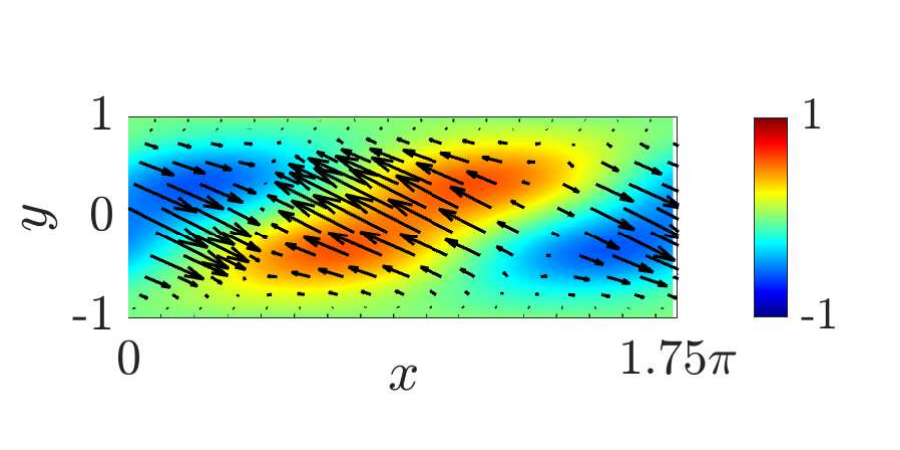}
         \caption{BT-LNT direct}     
    \end{subfigure}
    \vspace{\fill}
    \begin{subfigure}{0.45\textwidth}
         \includegraphics[width=\textwidth]{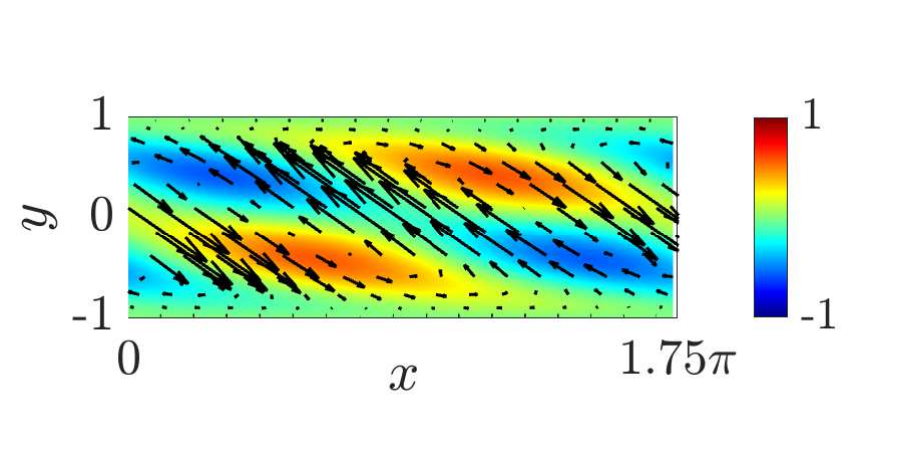}
         \caption{BT-LNT adjoint}
         
    \end{subfigure}
    \vspace{\fill}
    \begin{subfigure}{0.45\textwidth}
         \includegraphics[width=\textwidth]{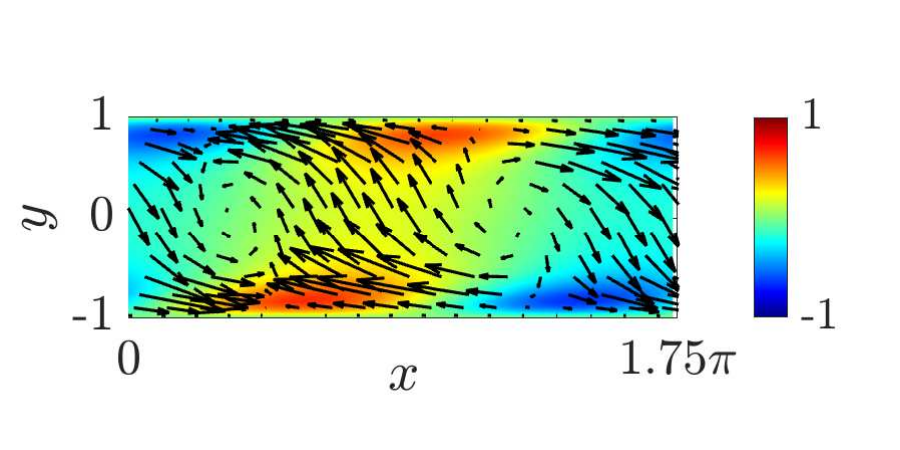}
         \caption{BT-LNTe direct}

    \end{subfigure}
    \begin{subfigure}{0.45\textwidth}
         \includegraphics[width=\textwidth]{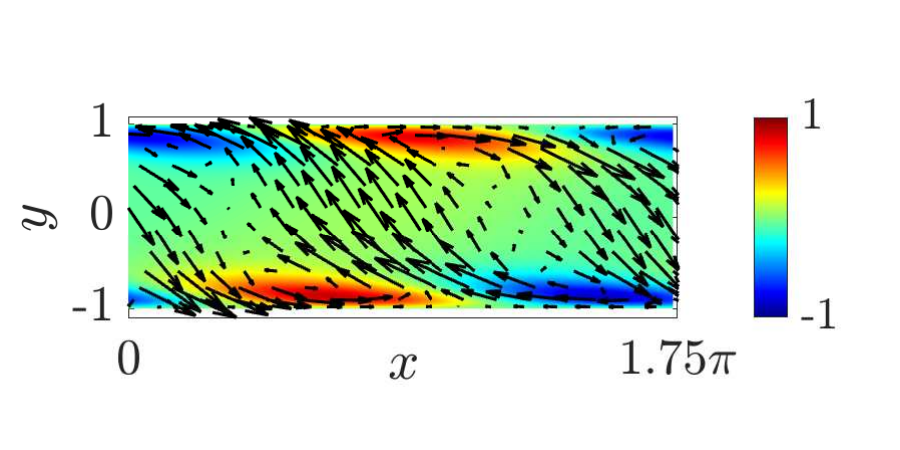}
         \caption{BT-LNTe adjoint}
    \end{subfigure}
    \caption{Cross-streamwise visualisation of the leading mode of different basis functions for ($n_x,n_z) = (1,1)$ on the $x$-$y$ plane. The positive/negative contours indicate positive/negative spanwise velocity, and the vectors represents the streamwise and wall-normal velocity components. Each mode is normalised by its energy, and the size of the vectors are scaled to be the same for all the cases.}
    \label{basis_visual_bt}
\end{figure}

The discussion above is summarised in Table \ref{pf_summary}. It is seen that there is a clear correlation between the states around which the ROMs perform well and those whose information was used to create the basis functions. The remaining part of the present study will be spent understanding where this correlation originates from a physical viewpoint. In particular, we shall focus on the dynamics around the laminar and turbulent state in relation to the flow structures of each mode, visualised in figures \ref{basis_visual} and \ref{basis_visual_bt} for $(n_x,n_z)=(0,1)$ and $(n_x,n_z)=(1,1)$, respectively.

\subsection{Laminar state}
In a previous study by \cite{Smith2005}, where a ROM was built using a Galerkin projection of POD modes obtained from turbulent state in plane Couette flow, it was reported that a sufficiently large number of $N_y$ is necessary for the ROM to have linear stability of the laminar base state. Appropriate capturing of the linear stability of the laminar state is crucial, especially from a dynamical systems viewpoint, because it is the most fundamental feature required to depict the correct state-space dynamics for the transition to turbulence in plane Couette flow: e.g. saddle-node bifurcation for the first non-trivial invariant solutions to the system \cite[]{Nagata1990}, the edge of turbulence \cite[]{Skufca2006}, and so on. For this reason, \cite{Smith2005} subsequently introduced a decoupling of each of the POD modes for $n_x=0$ or $n_z=0$, so that their streamwise and cross-streamwise components are assigned into two different orthogonal modes. This was to better capture the so-called `lift-up' effect \cite[especially for $n_x=0$ and $n_z\neq 0$; see also][]{Berkooz1991}, by which the streamwise component of velocity is amplified due to the energy from an interaction between mean shear and the cross-streamwise components. In doing so, they were able to restore the linear stability of the laminar state without considering large values of $N_y$. 

In the present study, we have shown that the C-LNL and BT-LNL cases retrieve the linear stability of the laminar state only with a single mode in the wall-normal direction ($N_y=1$). This was possible without introducing any decoupling procedure used in \cite{Smith2005}. We note that the basis functions for the C-LNL and BT-LNL cases are directly associated with the response of the given system around the laminar state to a small white-noise forcing: the C-LNL modes are physically the POD modes of the linear white-noise response of the laminar base state, while the BT-LNL modes combine the C-LNL modes with the observability modes that capture the flow structures responsible for the C-LNL modes. Therefore, they are expected to capture well the flow structures associated with the perturbation dynamics and the related stable manifolds of the laminar state. Despite this expectation, the related performance appears to be remarkably superior to that of the other basis functions, which require $N_y \gtrsim 20$ to achieve the linear stability of the laminar state. 

This is also evident in figure \ref{basis_visual}, where the leading modes used for the ROMs in the present study are visualised for $(n_x,n_z)=(0,1)$. The leading C-LNL (figure \ref{basis_visual}b) and BT-LNL modes (figure \ref{basis_visual}e) are structured with highly amplified streamwise velocity alternating in the spanwise direction (i.e. streaks), and they are highly energetic around the channel centre. However, the streaks captured in the leading POD mode have their largest amplitude near the wall, and they involve counter-rotating streamise vortices, the amplitude of which is considerably larger than that of the leading C-LNL (figure \ref{basis_visual}b) and BT-LNL modes. This implies that the most energetic modes relevant to turbulent state are quite different from the most energetic or dynamically relevant modes associated with the laminar base state even at the present Reynolds number ($Re=500$ or $Re_\tau=34$), which may well be viewed as low. Furthermore, the fact that the leading C-LNL and BT-LNL modes contain considerably smaller cross-streamwise velocity components relative to the POD mode supports the decoupling approach of \cite{Smith2005} employed to restore the linear stability of the laminar base state: the mode containing only the streamiwse component due to their decoupling would presumably be more appropriate to model the perturbation dynamics energetically dominated by the streamise velocity like the leading C-LNL and BT-LNL modes.  

\subsection{Balanced truncation modes revisited}
Controllability modes capture the most energetic flow structures from the response of the given linear system to a white noise forcing, whereas observability modes capture the structures leading to the most energetic flow response. As discussed in \S\ref{sec:2.3}, the balanced truncation approach considers a trade-off between controllability and observability of the given (linear) system, providing a ROM that can better describe the dynamical behaviour of the full system only with small DoF \cite[]{rowley_2}. This is also very well supported by the BT-LNL case, which performs better than any other ROMs around the laminar state (see \S\ref{sec:4.1}). Visualisation of the balanced truncation modes for ($n_x,n_z$)=(0,1) and $(n_x,n_z)=(1,1)$ in figures \ref{basis_visual} and \ref{basis_visual_bt} provides further physical insight into why the balanced truncation is so effective around the laminar case. For $(n_x,n_z)=(0,1)$, all the direct balanced truncation modes (the right eigenvector of $\boldsymbol{W}_c\boldsymbol{W}_o$), which were used as basis functions for the corresponding ROMs, are dominated by streamwise streaks (figures \ref{basis_visual}e,g,i), while the adjoint balanced truncation modes (the left eigenvector of $\boldsymbol{W}_c\boldsymbol{W}_o$) are mainly composed of counter-rotating streamwise vortices (figures \ref{basis_visual}f,h,j). This is reminiscent of previous studies where controllability and observability modes were shown to be dominated by streaks and counter-rotating streamwise vortices, respectively \cite[e.g.][]{10.1063/1.858894,HWANG_COSSU_2010_amplification,HWANG_COSSU_2010_1}, indicating that these balanced truncation modes contain physical elements to suitably capture the lift-up effect. It also explains well why the BT-LNL case captures the optimal transient growth so well compared to all the other ROMS in addition to the fact that the related modes are sampled from the dynamics around the laminar state. A similar conclusion can be drawn for $(n_x,n_z)=(1,1)$, illustrated in figure \ref{basis_visual_bt}. In this case, direct balanced truncation modes contain flow structures tilted downstream, while the corresponding adjoint modes have flow structures tilted upstream, indicating that they inherently contain suitable physical elements for the description of the Orr mechanism \cite[]{Orr1907,Jiao2021}, the energy production mechanism of spanwise vortical structures due to tilting by base flow shear.

Despite the substantial advantages in the description on the dynamics around the laminar base state, the ROMs based on balanced truncation (cases BT-LNL, BT-LNT, and BT-LNTe) were found to not perform substantially better than those based on controllability modes (cases C-LNL, C-LNT, and C-LNTe) for concerning the description of a turbulent state. In fact, for turbulent states, it was found that the POD case consistently outperforms all the other ROMs. As discussed above, the balanced truncation is able to account for the physical processes associated with the non-normality of the linearised Navier-Stokes equations, such as the lift-up effect and the Orr mechanism. It has been well understood that these processes are also highly relevant for the dynamics of coherent structures in turbulent flows: for example, the lift-up effect has been known to be the key process for the amplification of streaks \cite[e.g.][]{Hamilton_Kim_Waleffe_1995,bulter_farrel_93_pof,del_2006,HWANG_COSSU_2010_amplification}, and there has been considerable evidence on the importance of the Orr mechanism for coherent structures in turbulent flows \cite[]{Jimenez2013,Enciner2020}. Furthermore, in the case of BT-LNTe, the leading (direct) balanced truncation mode for the most energetic plane Fourier mode $(n_x,n_z)=(0,1)$, which represents the streaks, compares fairly well with the leading POD mode of turbulent flow (compare figure \ref{basis_visual}a with figure \ref{basis_visual}i) - the spatial structure of the streaks and the relative strength cross-streamwise velocity components compare much better than those of the BT-LNL and BT-LNT cases (for a further discussion, see also \S\ref{sec:5.3}). However, for $(n_x,n_z)=(1,1)$, the leading direct balanced truncation mode of the BT-LNTe case is seen to be considerably different from the leading POD mode of turbulent flow (figure \ref{basis_visual_bt}): the downstream tilt angle and the relative intensity of the spanwise velocity (figure \ref{basis_visual_bt}i), and the vector field indicating the wall-normal and spanwise velocity components of the BT-LNTe case are all seen to be considerably different from those of the POD mode (figure \ref{basis_visual_bt}a). In fact, in this case, the modes of the BT-LNL and BT-LNT cases are seen to compare better at least for the spanwise velocity, although their wall-normal velocity component also appears to be non-negligibly different from that of the POD mode. We note that the Fourier mode of $(n_x,n_z)=(1,1)$ has been understood as a consequence of streak instability and/or transient growth \cite[]{Hamilton_Kim_Waleffe_1995,schoppa2002}. In other words, the relevant physical process for this mode would be depicted by the linearised Navier-Stokes equations about a suitable streaky base flow, $\boldsymbol{U}_b(y,z)=\boldsymbol{U}(y)+\boldsymbol{u}_s(y,z)$, where $\boldsymbol{u}_s(y,z)$ represents the streaks (e.g. figure \ref{basis_visual}a). The leading POD mode for $(n_x,n_z)=(1,1)$ would then reflect the flow structures associated with the streak instability and/or transient growth, a nonlinear process that cannot be modelled by the linearised Navier-Stokes operator about turbulent mean flow $\boldsymbol{U}(y)$ with a simple white-noise forcing regardless of the presence of the eddy viscosity. This also explains why the POD case reproduces the qualitative behaviour of the correlation $C_{(0,1)}^{(1,1)}$ at low values of $N_y$ ($N_y=5$; figure \ref{fig:xcor}c), whereas all other ROMs are unable to do so even at $N_y=10$ (figure \ref{fig:xcor}f) - this correlation is associated with the phase relation between the streaks and the streak instability/transition growth.


\subsection{Linear operator modes with turbulent mean velocity and eddy viscosity}\label{sec:5.3}
Although the ROM based on POD modes shows the best performance in modelling turbulent state, the ROMs accounting for the information from turbulent state perform substantially better than those without it. This is particularly true for the C-LNTe and BT-LNTe cases, where the linear operator employs both the turbulent mean flow and the corresponding eddy viscosity. As briefly discussed earlier, the benefit of using such a linear operator for the generation of basis functions is most apparent for $(n_x,n_z)=(0,1)$. This Fourier mode depicts the streamwise elongated structures that have been understood to be well modelled by the linearised Navier-Stokes equations with turbulent mean flow and eddy viscosity \cite[]{HWANG_COSSU_2010_amplification}. In this case, the inertial part of the streamwise momentum equation is given by 
\begin{equation}
\frac{\partial \hat{u}}{\partial t} \sim -\hat{v}\frac{dU}{dy}, 
\end{equation}
implying that, for a given $\hat{v}$, $\hat{u}$ would be more energetic at which $dU/dy$ is large. This explains why the streamwise velocity of the leading modes from the linear operators with turbulent mean velocity tends to be relatively strong in the region close to the wall (C-LNT, C-LNTe, BT-LNT and BT-LNTe in figure \ref{basis_visual}).

\begin{figure}
    \centering
    \includegraphics[width=\linewidth]{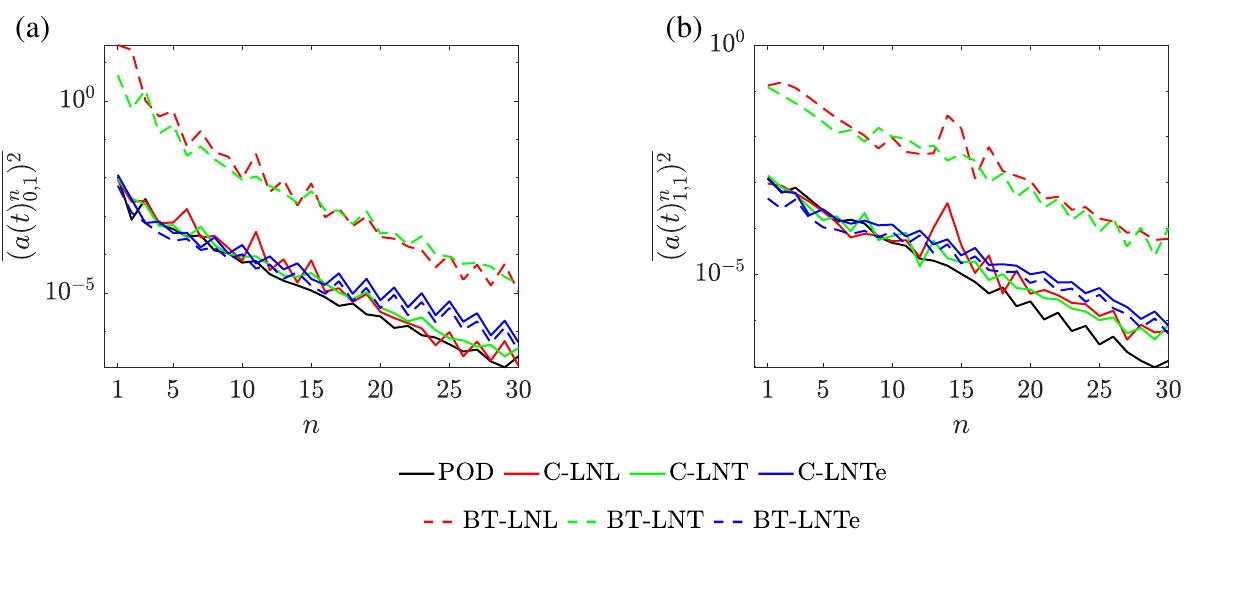}
    \caption{Time-averaged mode coefficient fluctuations: (a) $(n_x,n_z)=(0,1)$; (b) $(n_x,n_z)=(1,1)$.}\label{fig9}
\end{figure}

Furthermore, enhancing the linearised equations with the eddy viscosity model is seen to further improve the structure of the leading modes, making it more favourably comparable to the leading POD mode. The underlying reason can also be explained by the diffusion term of the eddy viscosity. In particular, the wall-normal transport of the streamwise Fourier component by this term is associated with the following part of the linearised Navier-Stokes equations: 
\begin{equation}\label{eq:5.1}
\frac{\partial \hat{u}}{\partial t} \sim \nu_t\frac{\partial^2 \hat{u}}{\partial y^2}+\frac{d \nu_t}{dy}\frac{\partial \hat{u}}{\partial y}. 
\end{equation}
Here, the first term on the right-hand side represents the diffusion by eddy viscosity, and the second term arises the nature of the eddy viscosity varying in the wall-normal direction. Given the nature of the first term that would enhance the diffusion and weaken the non-normality of the linearised Navier-Stokes operator, the related leading modes are expected to be more isotropic. This is consistent with the structure of the leading C-LNTe and BT-LNTe (direct) modes for $(n_x,n_z)=(0,1)$, which contain relatively large cross-streamwise velocity components (figures \ref{basis_visual}d,i) with amplitudes comparable to those of the leading POD modes (figure \ref{basis_visual}a). The second term on the right-hand side of (\ref{eq:5.1}) is given in the form of a wall-normal advection with a local speed of ${d \nu_t}/{dy}$, and this may be interpreted as a model for wall-normal turbulent (nonlinear) transport  \cite[see also][]{hwang_mesolayer_16,Ivan_eddyvisc_spod,Holford_Lee_Hwang_2024b}. Since $\nu_t$ vanishes on the wall, ${d \nu_t}/{dy}>0$ around the lower wall and ${d \nu_t}/{dy}<0$ around the upper wall. This term would then create an advection flux towards the wall, which is why the streamwise velocity of the leading C-LNTe and BT-LNTe (direct) modes is more energetic than that of the other cases near the wall, similar to that of the leading POD modes. 

Finally, it is worth mentioning that the balanced truncation modes of the linearised Navier-Stokes operator are non-orthogonal due to the advection term. Indeed, given the inner product defined for any solenoidal three-dimensional vector field in \S\ref{sec:background}, $\boldsymbol{B}^\dagger=\boldsymbol{C}$ \cite[$\boldsymbol{C}^\dagger=\boldsymbol{B}$, equivalently; see ][]{Holford2024}. Also, consider that $\boldsymbol{A}$ is the Stokes operator where the advection term is absent, $\boldsymbol{A}^\dagger=\boldsymbol{A}$. In such a case, $\boldsymbol{W}_c=\boldsymbol{W}_o$, which makes the controllability and observability modes identical. Consequently, for the Stokes operator, $\boldsymbol{W}_c\boldsymbol{W}_o$ becomes self-adjoint and the resulting balanced truncation modes are orthogonal. From this observation, in general, the linearised Navier-Stokes operator is expected to yield highly non-orthogonal balanced modes on increasing the Reynolds number. The use of these modes as the basis functions is then not necessarily robust for the construction of a ROM, as they generally do not guarantee rapid convergence to any arbitrary functions with increasing degree of freedom. 

To see this, the squared values of the mode coefficients for all ROMs are shown in figure \ref{fig9}. Indeed, except for the BT-LNTe case, where the corresponding linear operator employed the eddy viscosity model, the ROMs with balanced truncation modes (BT-LNL and BT-LNT) have mode coefficients an order of magnitude higher than those with orthogonal modes (i.e. POD and controllability modes). It is expected that these balanced-truncation-based ROMs will have even higher mode coefficients at higher Reynolds numbers, because of their increasing non-orthogonality, a potential issue in their applications to such flows. In this respect, adding an eddy viscosity model to the linearised Navier-Stokes operator appears to be useful, as it would reduce the non-orthogonal nature of the balanced truncation modes. This is well observed from the mode coefficients of the BT-LNTe case, where their values are comparable to those of the ROMs based on orthogonal modes.  


\section{Concluding remarks}\label{conclusion}
In the present study, we have investigated how the choice of basis functions affects the performance and convergence of Galerkin projection-based ROMs for Couette flow. To this end, POD modes from turbulent flow and controllability and balanced truncation modes of the linearised Navier-Stokes operator considering different base flows (laminar base flow and turbulent mean flow) and incorporation of an eddy viscosity model were considered. It was found that the performance and convergence of the ROMs constructed with various basis functions are significantly dependent on the state and are highly correlated with the information that each set of the basis functions contains. In the neighbourhood of the laminar base state, the balanced truncation modes from the Navier-Stokes operator linearised about the laminar base flow (BT-LNL) were found to be the best basis functions, as the resulting ROM well captures the linear stability of the laminar base flow and transient growth only with a very small degree of freedom ($N_y \simeq 5$). The controllability modes based on the same linear operator (C-LNL) were also effective in retrieving the linear stability of the laminar state and the related transient growth. In contrast, POD modes were found to be the basis functions to describe a turbulent state, and balanced and controllability modes based on the linearised Navier-Stokes operator employing turbulent mean flow and an eddy viscosity model (BT-LNTe and C-LNTe) were also effective for the construction of ROMs compared to other operator-based basis functions. These observations have also been discussed in relation to the dynamics and flow structures around the laminar and turbulent states. 

In a way, the conclusion here might have been expected somehow: the ROM utilising information from a state would converge quickly to model the dynamics around the same state. However, it is worth recalling that the degree to which each of the ROMs shows the state-dependent convergence appears to be remarkably high. For example, the cases BT-LNL and C-LNL only need a single mode in the $y$-direction ($N_y=1$) to retrieve the linear stability of the laminar base state, and BT-LNL can reliably reproduce the optimal transient growth only with $N_y=5$. Similarly, the POD case only needs $N_y=5$ to retrieve most of the statistical and dynamical characteristics of turbulent state, whereas all the other cases require at least $N_y=20$ to do the same. 

The ultimate goal of any ROM lies in the reduction of dimensionality of the original Navier-Stokes equations. However, in practice, it is very difficult to achieve this goal without losing any of the modelling capacity of the ROM. At high Reynolds numbers, the expected dimension of any ROMs would have to be non-negligibly large, as it would also increase with the Reynolds number: for example, the Lyapunov (Kaplan-Yorke) dimension of the present minimal Couette flow at a slightly lower friction Reynolds number ($Re_\tau=34$) is $D_L=14.8$ \cite[]{Inubushi2015}, but a similar setting in Poiseulle flow at higher Reynolds number at $Re_\tau=80$ is $D_L \sim 780$ \cite[]{Keefe1992}, indicating a very rapid increase of the Lyapunov dimension with the Reynolds number. This also implies that the dimension of the ROM would also need to increase very rapidly, especially if one wishes to capture the turbulent dynamics very accurately. It is then practical and probably necessary that a ROM for high Reynolds numbers would need to make some sacrifice in its modelling capability to serve its original purpose. This rationale is similar to that of large eddy simulation (LES), where the dynamics of small scales are replaced with a simple subgrid-scale model or are not even modelled (e.g. implicit LES). The observation in the present study is useful in this regard. For example, at high Reynolds numbers, the laminar base flow is practically unseen because of its very small basin of attraction or its strongly unstable nature. Then, as long as a ROM is able to describe turbulent state well, there would probably be no need for this ROM to capture the dynamics around the laminar base flow accurately. From this viewpoint, the present study highlights the usefulness of POD modes from turbulent flows.These POD modes also need to be obtained from simulation and experiment data. However, obtaining large number of converged modes requires, apart from a significant amount of temporal data, a significantly large discretisation. This can be particularly challenging for experimental data. The balanced truncation and controllability modes from the linearised Navier-Stokes equations with turbulent mean flow and an eddy viscosity model may be a suitable alternative to the POD modes (especially at high Reynolds number), despite their limitations. Developing a ROM with these modes at high Reynolds numbers is our on-going task combining with a suitable data-driven augmentation \cite[e.g.][]{Chua_Khoo_Chan_Hwang_2022}.

\backsection[Acknowledgements]{ }
This work was initiated at the Isaac Newton Institute for Mathematical Sciences at the University of Cambridge through the programme ‘Mathematical aspects of turbulence: where do we stand?’ supported by the Engineering and Physical Sciences Research Council (EPSRC; EP/R014604/1). 

\backsection[Funding]{ }
Y.H. gratefully acknowledges a financial support from CAPES (Brazilian Federal Agency for Support and Evaluation of Graduate Education) for the visit to Instituto Tecnológico de Aeronáutica. 

\backsection[Declaration of interests]{ }
The authors report no conflict of interest.



\appendix




\bibliographystyle{jfm}
\bibliography{jfm}




\end{document}